\documentclass[a4paper,fleqn,usenatbib]{mnras}

\usepackage{newtxtext,newtxmath}

\usepackage[T1]{fontenc}
\usepackage{ae,aecompl}
\usepackage{todonotes}

\usepackage{graphicx}	
\usepackage{amsmath}	
\usepackage{amssymb}	

\title[Nova or Supernova? Revisiting guest stars]{Counterparts of Far Eastern Guest Stars: Novae, supernovae, or something else?}

\author[Hoffmann et al.]{
Hoffmann, Susanne M.,$^{1}$\thanks{E-mail: susanne.hoffmann@uni-jena.de (PAF, FSU)}
Vogt, Nikolaus,$^{2}$
\\
$^{1}$Physikalisch-Astronomische Fakult\"at, Friedrich-Schiller-Universit\"at Jena, Germany\\
$^{2}$Instituto de Física y Astronomía, Universidad de Valparaíso, Chile\\
}

\date{Accepted 2020 May 28. Received 2020 May 20; in original form 2020 April 28}

\pubyear{2020}

\begin{document}
\label{firstpage}
\pagerange{\pageref{firstpage}--\pageref{lastpage}}
\maketitle

\begin{abstract}
 Historical observations of transients are crucial for studies of their long-term evolution. This paper forms part of a series of papers in which we develop methods for the analysis of ancient data of transient events and their usability in modern science. Prior research on this subject by other authors has focused on looking for historical supernovae and our earlier work focused on cataclysmic binaries as classical novae. In this study we consider planetary nebulae, symbiotic stars, supernova remnants and pulsars in the search fields of our test sample. We present the possibilities for these object types to flare up visually, give a global overview on their distribution and discuss the objects in our search fields individually. To summarise our results, we provide a table of the most likely identifications of the historical sightings in our test sample and outline our method in order to apply it to further historical records in future works. Highlights of our results include a re-interpretation of two separate sightings as one supernova observation from May 667 to June 668 CE, the remnant of which could possibly be SNR G160.9+02.6. We also suggest the recurrent nova U Sco as a candidate for the appearance observed between Scorpius and Ophiuchus in 891, which could point towards a long-term variability of eruption amplitudes. In addition, we find that the `shiny bright' sighting in 1431 can be linked to the symbiotic binary KT Eri, which erupted as a naked eye classical nova in 2009.  
\end{abstract}




\section{Introduction}
This is the fifth and last paper of our series developing a new method of analysing the historical sightings of transients by using the many electronic databases developed during the past few decades. In our first two papers, \citet{vogt2019} and \citet{hoffmann2019}, we reasoned for this study by expecting some dozen or some hundred additional data points in the 2.5 millennia. In order to provide these additional data points, \citep{hoffmannVogtProtte} described a new method to translate the positions described in historical text in search fields usable with modern databases. This method was applied to a test sample of 25 cases selected $\sim185$ \citep[Tab.~2 and 3]{hoffmannVogtProtte} reports. In \citet{hovoMNRAS2020} we probed and analysed the CVs in these fields as counterparts of potential historical novae. As in historical cases only an appearance or disappearance is reported, the nature of the event is never certain. Therefore, with this paper we've come full circle by analysing the alternative counterparts in case the certain event was not caused by an eruption in a cataclysmic variable: We query databases for alternative object classes able to flare up to naked eye visibility -- other types of novae and other types of stellar transients. 

 \subsection{Previous work} 
  Earlier efforts to locate and determine the nature of observations in historical records have most industriously and famously been provided by F. Richard Stephenson \citep{stephenson,steph77,stephGreen,greenSteph2003,greenSteph2017} and his colleagues, e.\,g. \citet{green2019,yau}, with the focus of supernovae. Other scholars used Stephenson's results for further studies, e.\,g. \citet{duerbeck,shara2017_ATcnc_steph,hoffmann2019} on classical novae or \citet{fujiwara2003,fujiwara2005,hamacher2010,hamacher2018} on the long-term variability of stars. \citet{hoffmann2019} pointed out some common misunderstandings of Stephenson's work, especially the interpretation of the positions in his catalogues as point coordinates instead of centres of areas. To these areas, we applied error circles with a radius of 4\degr\ derived from known historical supernova identifications \citep[Tab.\,6]{hoffmannVogtProtte}. By re-interpreting the historical texts we also added some further records from Stephenson's predecessors in the efforts of collecting records of transients and extracting catalogues of diverse types from them \citep{hsi,ho,xi+po,pskovskii}. 

 In a preliminary study, \citet[Tab.\,3 and 4]{vogt2019} estimated the amount of classical novae we expect to be observed within 2.5 millennia (the range of somehow reliable historical text records) and kindled a study of the historical records with a focus on the life of cataclysmic binaries after their eruption \citep{tappert2012}. This was performed by \citet{hoffmannVogtProtte} investigating the questions of naked eye visibility (Fig.\,3 therein and \citet{protteHoffmann2020}) of classical novae and defining search fields (Tab.~2 and 3 therein) for the modern counterparts of historical transients. 

 \subsection{Motivation}
  Historical observations of transient phenomena may help to develop models of evolution of stellar systems -- close binaries with white dwarfs \citep{shara1984,pat2013,miszalski2016,duerbeck,shara2017_nov1437}, neutron stars \citep{schlier,mayall+oort,baade,morgan2007}, or other systems with two components in a common envelope or any type of strong eruption. They also could help to understand the evolution of planetary nebulae, nova shells, and supernova remnants: The age determinations of these objects are rather uncertain. A historical observation which enables us dating the appearance within a month or within a certain `reign period' (a couple of a few years) would narrow the observational error bars, usually being in the order of decades to millennia, cf. e.\,g. \citep{reynolds2019,stafford2020} and the age ranges given in the catalogue of supernova remnants \citep{manitoba}. 
 
  Summarizing, there is a strong wish to enlarge the temporal range of usable observations on the one hand but huge uncertainties and difficulties in interpreting the few historical data points on the other hand. Nevertheless, \citet{hovoMNRAS2020} investigated the search fields derived from text interpretation to determine if a CV could produce a naked-eye nova. In most cases, we could find candidates.  
  
 \subsection{Difficulties and open questions} 
 \textbf{From the historical point of view,} there are the following difficulties \citep{hoffmann2019}:
 \begin{itemize}
 \item Historical observers usually did not understand what they saw. So they did not apply our scientific standards in their reporting and follow-up observing.
 \item Many historical observations are lost, so we do neither have a complete set of observations nor a set of astronomical diaries (except from Babylon) and cannot reconstruct light curves of the historical sightings.  
 \item Most of the preserved records are not preserved in scientific context but in copies of copies of copies of interpretation and selection by chroniclers.
 \item The vocabulary which was common to describe an observation changed many times during history. Thus, the exact terminology applied in a historical record is ($i$) not necessarily the original one (because the chronicler some centuries later might have used a different term than the original record of observation) and ($ii$) not necessarily the one we would use today. 
 \item Some historical `science cultures' did not report transient phenomena at all or not in astronomical context because they believed them to be something like weather. Thus, the preserved collection of such records is far from complete -- but the absence of observation is not the observation of absence of a certain phenomenon (e.\,g. eruptions). 
 \item The biggest text corpus of records of transients is preserved from the Far East because the Chinese astral science and derivative astral science cultures, e.\,g. Korea, Japan, and other former colonies of China, served the belief that transients indicate something of political importance: The Chinese asterisms are considered as a projection of all areas of political and social life into the sky (there are, e.\,g. the asterism of the Celestial Market Place representing the people or the asterism of the Forbidden Palace with the houses of the emperor, the empress, the crown prince, the other children, the maids and servants, the highest officers and ministers of the state). The transients -- no matter of their astrophysical nature -- were believed to indicate that something (good or bad) happens or will happen to the group which is represented by the asterism in which it appeared. Thus, there was a strong focus for the astronomical observers to look for and report transients. 
 \item In many cases the position is given very imprecise: For the divine purpose, mainly the constellation was important and not a certain star with known coordinates. Luckily, many Chinese constellations are much smaller than the IAU-constellations. Some only consist of two or three stars -- but still there are some really huge ones, e.\,g. the asterism of the Forbidden Palace covers almost the whole circumpolar region.  
 \item  An observation by astronomers of different cultures who usually use different constellations could help to improve the positioning of the observation by using the intersection of the reported constellations \citep{neuhaeuserKomet}. In many other science cultures, the `research focus' was different and was triggered by other religious beliefs. Thus, a Far Eastern record is often not confirmed by an observation in other cultures. 
 \item The chronicles do not preserve all transients but only those which later turned out to predict something correctly and which could, therefore, be used to narrate the (hi)story.  
 \end{itemize} 
 
 \textbf{From the astronomical point of view:} Due to the lack of historical follow-up observations of appearances, even in the cases with only few counterpart candidates, it is not certain that the transient really was a classical nova and not a supernova or something else. Due to the brevity of the historical records in our sample list \citep{hoffmannVogtProtte} and the resulting inability to reconstruct (or at least estimate) light curves, the nature of an event remains always uncertain. We already excluded the currently known Mira stars because of their faintness \citep{vogt2019} but all other possibilities need to be checked.

 Thus, the present study investigates alternatives: Could it be that an event of our shortlist possibly refers to a supernova instead of a nova? If we did not yet find a best candidate CV: Is there possibly a planetary nebula in the field with a close binary in its centre? Could one of the planetary nebulae in the field be misclassified and turn out to be a nova shell? The genesis of a planetary nebula normally is a long process and not accompanied by a brightening of the system but are there other types of evolution in such a system which could cause a sudden appearance and disappearance? Are there symbiotic binaries (Z~And-type) which could have erupted as classical nova?

\section{Candidate Types}
This section is dedicated to summarizing the relevant qualities of symbiotic binaries, PNe, PSRs, SNRs, X-ray binaries that are important for our study. Additional object classes which could be relevant to explain historical guest stars are still missing: Currently, we do not regard variables with small amplitudes, Be-flares or with unpredictable behaviour and we do not yet probe for possible microlensing events.

  
  \subsection{Planetary nebulae} 
  There are two reasons for regarding PNe: ($i$) The nebula could be misclassified as PN and indeed be an SNR (as in case of the Kepler SNR) or nova shell (as in case of PN Te~11). ($ii$) Even a planetary nebula or its central star could have flared up but for longer timescales than considered here: There are rebirth scenarios for PNe as observed for Abell~30 and diffusion induced novae (DIN) like suggested for CK Vul, the Nova~1670, \citep{millerBertolami2011} before \citet{kaminski2015} came up with the hypothesis of a `red transient', i.\,e. the remnant of a star merger.\footnote{Nova 1670 Vul was observed and described by several European astronomers including th most famous observations by Hevelius in Poland. They describe strong variability for roughly two years before the object faded away from view completely. This is an extraordinarily well described sighting and seems to be explained, which is why it is not included in the scope of our paper.}
  
  \subsubsection{PNe as indicators for historical flares}
  The genesis of a PN is normally a long process over millennia and millions of years and not accompanied by a single short-term flare of the star. Thus, a definite PN is rather uninteresting for the study of historical transients. The estimate of the kinematic age of a nebula only indicates when the process started. However, the complex structure of planetary nebulae includes layers of younger events. If parts of it result from processes only some thousand years later, i.\,e. in historical epochs, one could speculate that there might have been an additional eruption or a rebirth \citep{bloecker2003} which had been observed as a guest star. Particularly interesting are processes which could cause a brightening of the compact central stars (CSPN): CSPN could be white dwarfs (WDs), binaries containing a WD, or Wolf-Rayet stars (WR). The majority of CSPN Wolf-Rayet stars are of nitrogen type (WN) but there are also cases of the carbon type (WC).

 In some rare cases, central stars of PNe experience a very short second life when nuclear fusion is kindled again in the shell of the hot pre-white dwarf star (He-shell flash) \citep{falk2006, todt2015}. These rebirth scenarios evolve on timescales of years. They are, thus, in principle interesting for the search for counterparts of historical `new stars' such as Nova 1670~Vul and the events without given duration but not in the cases of our selection with given durations of only several days.
   
 The remaining option to flare up a CSPN would be diffusion induced novae (DIN) \citep{millerB2011} which could possibly explain the observed properties of [WN/WC]-central stars of PNe. DIN are rare cases where the central star experiences a very late thermal pulses (VLTP), properly known only for a handful of events (V605 Aql, V4334 Sgr, NSV~11749): Initially, they increased in brightness more than 3.5~mag in V, then showed a steady stage of brightness at maximum and then completely faded away from view rather quickly with a R~CrB-like decline and carbon-dust emission. However, this evolution takes place on the timescales of a few years (observed cases: 6.3 and 4.4~years) and is, thus, also not urgently relevant for those of our selected guest star events with a given short duration. 
  
  \subsubsection{Examples and criteria for misclassified PNe}
 Circularly shaped nebulae are often classified as `planetary nebulae' at first glance and only further investigations show the true nature of the object. The most famous case is probably the remnant of Kepler's supernova: It had been enrolled in the Catalogue of Galactic planetary nebulae \citep{kohoutek2001} before it was unveiled as `not PN' \citep{frew2013} but it is also suggested as `not SNR' \citep{acero2016}. 

  Further, there is relatively high likelihood for a misclassification of a nova shell as planetary nebula (PN) \citep{miszalski2016,frew2013,shara2017_nov1437}. This is caused by the high similarity of these types of nebulae. Both can have a central stellar remnant and show similar chemical abundances. They even show the same forbidden lines in the spectrum \citep{kwok2007}. 
  
 The only criterion to distinguish between the two categories seems to be the expansion rate because the initial momentum at nova eruptions (thermonuclear runaway on the surface of a white dwarf) is larger than the initial momentum of the red giant thermal pulsation driven ejecta or the final repulsion of outer shells of an old star as planetary nebula. Therefore, the average expansion rates of nova shells should be higher than for PNe by a factor of 10 to 30. However, all physical quantities are observed in distributions, most expansions are decelerated with time, and the average value can only give an impression or suggest a tendency and is never a final proof. A fast planetary nebula and a slow nova shell could be exchanged; it will always need a closer inspection of the individual objects. In our study, we aim to lay the foundation for further observational research.    
 
 \subsection{Apparent pairs of nebulae and close binaries} 
 \subsubsection{Fraction of PNe with central binaries.} According to \citep[p.\,452]{warner1995}, $\sim13$\,\%\ of PN are known to have binary nuclei and at least 17\,\% of late F and G stars are expected to undergo a common envelope evolution. Newer results estimate much larger binary fraction of CSPN of $60-80\,\%$ \citep[p.\,72]{Boffin2019}. Given a fraction of binarity for main-sequence progenitors of $(50\pm4)$\,\%\ and their observed sample of 35~CSPNe \citet{douchin2015} derive a fraction of $(40\pm20)$ to $(62\pm30)$\,\%\ in $I$ and $J$ band, respectively \citep{douchin2015PhD}. As the fraction of binaries is independent from our observational filter, we use a binary fraction of CSPNe of $62$\,\%\ because many of the $I$ band binaries could also be visible in the lower energy $J$ band. They point out that including white dwarfs and other `evolved, hot companions, which may constitute up to a quarter of all companions,' might increase the binary fractions by 13 and 21 points respectively. This is still compatible with the summary in the book \citet[chap.\,6.1]{Boffin2019}. For our study, this means that we expect to find 13 to 21\,\%\ or up to a quarter of the planetary nebulae with central star white dwarf. 

  \begin{table}
 	\caption{Proper motion (pm) and parallax (plx) of the brightest known classical novae ($V_\textrm{max}\leq4$~mag). The proper motion is the sum of the two components in right ascension and declination. The table shows that these objects at average distance of 1,042~pc move less than a minute of arc within the given timespan of 2,500~years.}
	\label{tab:pmCN}
 \begin{tabular}{lrrr}
 \hline
      & plx & pm  &  in 2500 years \\
 Nova & in mas &mas/year  &  arcsec\\
 \hline
V603 Aql  &  3.19  &   14.77 &  36.9 \\
V1369 Cen &  3.64 & 20.71 &  51.8\\
V1500 Cyg  &  0.777 & 8.45  &  21.1\\
HR Del      &  1.04  &  9.47  &  23.7\\
DN Gem  &  0.729   &  4.39  &  11.0\\
DQ Her    &   2.00  & 12.46   & 31.1\\
V446 Her   &  0.744  &  6.76  &   16.9 \\
V533 Her  &  0.83   &   1.67  &  4.2\\
CP Lac   &  0.86   &   8.64  &   21.6\\
GK Per &   2.26   &  18.46  &  46.1\\
RR Pic  &   1.95   &   4.78 &  12.0\\
CP Pup    &  1.23 &  3.00  &  7.5\\
V598 Pup &  0.47  &  7.32 &  18.3\\
 \end{tabular} 
 \end{table}
 \subsubsection{Search for misclassified PNe.} In case of wrongly categorised nova shells, we expect to find a CV within or in close vicinity to the gaseous nebula which is classified as PN. For our search for explanations of historical sightings, we are interested only in those CVs which could have caused an eruption to naked eye visibility. As the temporal range of the observation of guest stars is only $\sim2,500$~years, the CV which caused the nova remnant should not have a big distance to the centre of the shell. The angular separation of the CV and the shell centre depends on the difference of their proper motions. The maximum proper motion known of any star (Barnard's star) takes the object 7\fdg2 in 2,500~years but as the nova shell originates from the moving star the difference of their proper motion is only the (added or subtracted) component of the CV's proper motion change due to the eruption and the accompanying repulsion of material. This difference in proper motion will separate the CV and its shell only a few seconds of arc within 2,500~years or in rare cases maybe a few minutes of arc. Tab.~\ref{tab:pmCN} lists the proper motions of the brightest known novae and the maximum value of a shift of $51\farcs8$ within 2,500 years; the difference to the motion of the nova shell would be even smaller. As the proper motion of a star does not depend on the stellar type and eruption behaviour, this table only represents a random subset of stars. The table gives an impression that small angular separations between PN and CV can be applied as our selection criterion. 

However, as we are neither certain that our historical sightings were caused by classical novae nor that the causing object's distance is at a distance of the order of typical known nova parallaxes, we decided that our method should probe the field of 1\degr\ around each PN centre for CVs, XBs, and symbiotic binaries. In a second step, we, then, selected subsets of smaller angular separations, starting with the selection of apparent pairs closer than $10\arcmin$ for the discussion in this paper. 

 \subsection{Symbiotic stars}\label{chap:methZAND}
 When the first symbiotic binaries were discovered in the 1930s (AX~Per and CI~Cyg), their spectra were puzzling scholars because they show features of most different types: Their emission lines were similar to those of PN and they exhibit TiO bands like only the very cool M giants and He\,{\sc ii} (4686) which was known as typical for very hot O-type stars \citep{munari2019}. This featuring explains that nebulae of symbiotic stars have a high potential to be initially misclassified as PN and that their central stars are binaries. 
 
 According to \citet[p.\,239]{bode} symbiotic novae are the direct link between cataclysmic variables and symbiotic stars; they originate in cataclysmic binaries with a red giant as donor \citep{darnley2019}. The symbiotic stars of the so-called Z~And type comprise an inhomogeneous group of close binaries consisting on a compact component (normally a white dwarf or neutron star) and a late-type giant or super-giant. Their orbital periods are of the order of $2-3$ years and their light curves show irregular, sometimes  Mira-like variability with amplitudes of $\sim4$~mag at time scales similar to their orbital period. A symbiotic star primary accretes material not only by Roche lobe overflow but also by strong wind emission from the red giant star. A recent review on symbiotic stars is given by \citet{munari2019}.

 As in symbiotic stars the donor, a red giant, could cause a rather large mass transfer rate, nova eruptions could happen more frequently, as in the recurrent novae. Indeed, there are striking similarities between symbiotic stars and recurrent novae (Nr), i.\,e. novae with more than one single eruption observed. In a recent review on Nr type novae \citep{darnley2019} subdivides them into three classes: 
 \begin{enumerate}
 \item the RS~Oph class (sometimes also called T~CrB class \citep{warner1995}) with $P_\textrm{orb} = 225 - 520$~d,
 \item the U~Sco class with $P_\textrm{orb}\sim1$~day,
 \item and two known cases with even much smaller  $P_\textrm{orb}\sim0.1$~d, called T~Pyx class by \citet{warner1995}.
 \end{enumerate}  
 The latter two belong to the CVs, while the members of the RS~Oph class contain red giant secondaries, the same configuration as the Z~And class. In addition to the class Nr, there are also some classical novae (classified as Na, Nb and Nc according to their speed of decline) which might be related to the symbiotic stars: The VSX catalogue lists a total of 13 novae with orbital periods between 192 and 1011~d, five of them are recurrent novae (T~CrB, V407~Cyg, RS~Oph, V3890 Sgr and V745 Sco), three are rapidly declining novae of Na-type (KT~Eri, GR~Sgr and V1187 Sco) and the remaining five cases are extremely slowly declining novae of class Nc (V1329 Cyg, V2110 Oph, HM~Sge, ASAS J174600-2321.3 and PU~Vul). 

 The average total eruption amplitudes of Nr novae is 9.5~mag, of Na novae 9.8~mag and that of Nc novae only 7.2~mag (with extrema between 5.0~mag and 11.2~mag considering all 13 cases). All these amplitudes are significantly smaller than those of most of the classical novae with amplitudes of up to 13~mag, cf. \citet[Fig.\,1]{vogt2019}.  

 It should be emphasised that three of the above-mentioned novae had been naked eye objects: T~CrB (2.0~mag), RS~Oph (4.3~mag) and KT~Eri (5.4~mag). Therefore, it seems to be quite possible that some of the objects nowadays catalogued as Z~And stars had generated a nova eruption during the past 2,500 years.


 \subsection{X-ray binaries} 
 There are three main types of X-ray binaries: low mass X-ray binaries (LMXB), intermediate-mass X-ray binaries (IMXB), and high mass X-ray binaries (HMXB). These systems are binaries with mass transfer between the partners where the accretor is a compact object, i.\,e. neutron star, black hole, or white dwarf. The donor can be of O-, B, Be-type or supergiant for a HMXB, of A- or F-type for the rare IMXB and of dwarf star, white dwarf, or evolved star for LMXB. The source of the X-ray emission can be the accretion disk or the surface of the compact object. Occasionally, point sources like Seyfert galaxies and quasars are (accidentally) listed among the HMXB and will be found with our query.

  Initially, we considered mainly the subset of XBs which could erupt as classical novae \citep{hovoMNRAS2020}. In this paper, looking for all possibilities of flaring up, we include all types of X-ray binaries in the initial query and, then, filter those of them which are able to brighten in visual wavebands. Accreting black holes, e.\,g. could show flares also in optical pass bands like in the cases of the double-back hole blazar OJ287 listed as X-ray source in the catalogue \citep{massaro2009} but of which optical observations are known since 1900 \citep{hudec2013} or the HMXB with black hole V4641 Sgr (AAVSO light curve peaks at 12.8~mag, VSX even enrolls a report of 9.1~mag in V). The existence of stellar black holes in naked eye multiple star systems has been proven by the discovery of the not accreting stellar BH in the naked eye triple system QV~Tel \citep{rivinius2020}, $\sim5.3$~mag in quiescence. These examples demonstrate the optical detectability of X-ray sources and black holes.

 \subsection{Supernova remnants and pulsars}\label{chap:snrMeth}
 Additionally, we checked our search circles for supernova remnants (SNRs) or pulsars (PSRs). Out of the eight known historical supernovae which we used as comparison in our earlier papers, only three are associated with pulsars (Tab.~\ref{tab:histSN}): SN~1054, SN~1181, and SN~386 are matched (or suggested to match) with the pulsars PSR~J0534+2200, PSR~J0205+6449, and PSR~J1811-1925, respectively. These pulsars have spin periods of 33.392~ms, 65.716~ms, and 64.667~ms and typical slowing rates leading to characteristic ages $\tau_c$ of $1.26\cdot10^3$, $5.37\cdot10^3$, and $2.33\cdot10^4$~years, respectively. 
\begin{table*}
	\centering
	\caption{The currently known historical supernova remnants, the likelihood of their identification with the given event, their pulsars with their periods $P$ and characteristic ages $\tau_c$.}
	\label{tab:histSN}
	\begin{tabular}{cllllrr} 
		\hline
 	&\multicolumn{3}{c}{U Manitoba cat.} &\multicolumn{3}{c}{ATNF cat.}\\
		year	&SNR & 	 &PSR  &	$P$/s	&$\dot{P}$ 	&$\tau_c$/yr \\
		\hline
185		&G315.4-02.3 	&probable	&--			&&& \\
386		&G011.2-00.3 	&suggested	&PSR J1811-1925	&0.064667	&$4.4\cdot10^{-14}$	&$2.33\cdot10^4$\\
393		&G347.3-00.5	&probable	&--			&&&\\
1006	&G327.6+14.6	&certain	&--			&&&\\
1054	&G184.6-05.8	&certain	&PSR J0534+2200 &0.0333924123	&$4.21\cdot10^{-13}$	&$1.26\cdot10^3$\\
1181	&G130.7+03.1	&probable	&PSR J0205+6449 &0.06571592849324	&$1.94\cdot10^{-13}$	&$5.37\cdot10^3$\\
1572	&G120.1+01.4	&certain	&--			&&&\\
1604	&G004.5+06.8	&certain	&-- 		&&&\\
		\hline
	\end{tabular}
\end{table*}

 To derive the characteristic age $\tau_c=P/ 2\dot{P}$, it is assumed that the pulsar spins down by pure magnetic dipole braking and that the initial spin period is negligible compared to the present period. Therefore, the characteristic age usually overestimates the true age. With regard to the numbers in Tab.~\ref{tab:histSN} it is clear that $\tau_c$ gives only the correct order of magnitude, i.\,e. $10^3$ to $10^4$ years for supernovae in the past 2 millennia.
 
 However, in many cases the deceleration rate and, thus, the characteristical age $\tau_c$ of a PSR is unknown. Based on our current knowledge of PSR evolution, we can only judge that PSRs with long periods $P$ (seconds) and $\tau_c$ of $10^5-10^7$ years are highly unlikely but we cannot exclude them.
 
 Only the millisecond pulsars (MSP) with $\tau_c$ of the order of $10^7$ to $10^{10}$~years can be excluded. They are `reborn' pulsars that spun up due to accretion for many centuries or millennia. Thus, they are certainly too old for our short historical timespan which is not enough time for a pulsar to accrete enough material to experience a rebirth. MSPs are characterised by an extraordinarily short period $<16$~ms \citep{thorsett1993,nice1995,ray1996,hobbs2004} and a very modest change of this fast rotation which leads to the high characteristic age $>10^7$~years. 
 
 Our careful revision of all objects found in our search circles will, therefore, use the characteristic age of pulsars as an upper limit but there is no proper criterion to decide whether or not a certain pulsar might have caused a historical event; it is only a likelihood.  

 \section{Method} 
 \subsection{Workflow} 
 There are several catalogues for planetary nebulae, e.\,g. \citet{perek1967,allen1985,kohoutek2001,mash2006,frew2013}, several catalogues of supernova remnants, e.\,g. \citet{green1988,manitoba,anderson2017,green2019}, catalogues of pulsars, e.\,g. \citet{hulse1974,thorsett1993,ray1996,atnf}. Some of them enroll only a certain type of these objects, e.\,g. only young or only old PN, only millisecond pulsars or only those discovered by a certain instrument, only Galactic SNRs or only those discovered in radio band. In order to be not in any way biased by such a pre-selection and get the maximum amount of possible objects of the wanted type, we chose to extract our lists from databases (VSX, Simbad) which are known to output much more than needed, many suspicious and questionable objects of any type and even old classifications. This way, the chance of missing a possible target is minimised but a careful manual consideration of the output is required. 
 
 \textbf{Step 1:} We extracted the current list of known planetary nebulae (PNe) and PN candidates, supernova remnants, and pulsars from the CDS Simbad database \citep{wenger2000}, the current list of known cataclysmic variables (CV), X-ray binaries (XB), and symbiotic stars (ZAND) from the VSX database of the American Association of Variable Star Observers \citep{watson}. We used the query for \textit{otype} (object type) 'PN' and 'PN candidate' in CDS Simbad to obtain the current catalogue of planetary nebulae and \textit{otype}$=$'SNR' and 'PSR' for supernova remnants and pulsars. There is a huge number of extraGalactic SNRs, of misclassified PNe and PN candidates which has to be filtered in a later step when considering the objects in our search fields in more detail and study the literature on them. In known cases of misclassification there is a note in Simbad that the object is considered `not a PN'. We found a total of 23,188 planetary nebulae (including the candidates), 2,904 PSRs, 2,120 SNRs, 12,889 CVs, 186 XBs (including both, low-mass and high-mass XBs), and 286 symbiotic stars of Z~And type. 

 \textbf{Step 2:} In the next step, we mapped the full lists of objects into our star charts from \citet{hoffmannVogtProtte,hovoMNRAS2020} and found plenty of objects in each of our search fields (see Online-Only Appendices for the maps). The manual genesis of Tab.~\ref{tab:pn} and \ref{tab:snr+psr} from our interactive maps is a more sophisticated method than the earlier described VSX-probing with search circles \citep{hovoMNRAS2020}. This method also applies for queries in other databases and it allows to have the same perspective as the ancient astronomer describing the position of the considered object in one or another asterism. However, the search circle coordinates as provided in \citet{hoffmannVogtProtte} still serve for a quick reproduction.

 \textbf{Step 3:} Searching close apparent pairs of PNe and common envelope binaries, we first computed all angular separations of all these pairs in the sky. Second, for the close pairs in our search fields we did a literature study, tried to find out the spacial distances in order to identify possible real pairs. 
  \begin{figure}
    \caption{Flowchart of the process of evaluation: The input of `neb' is the joined table of `PN' and `PN candidates' in Simbad and `stars' is the joined table of all CVs, X-ray binaries and symbiotic stars currently registered in the VSX. $\zeta$ is the orthodrome as defined in  Equation~\ref{eq:dist}.}
    \label{fig:flow}
    {\centering 
	\includegraphics[width=.62\columnwidth]{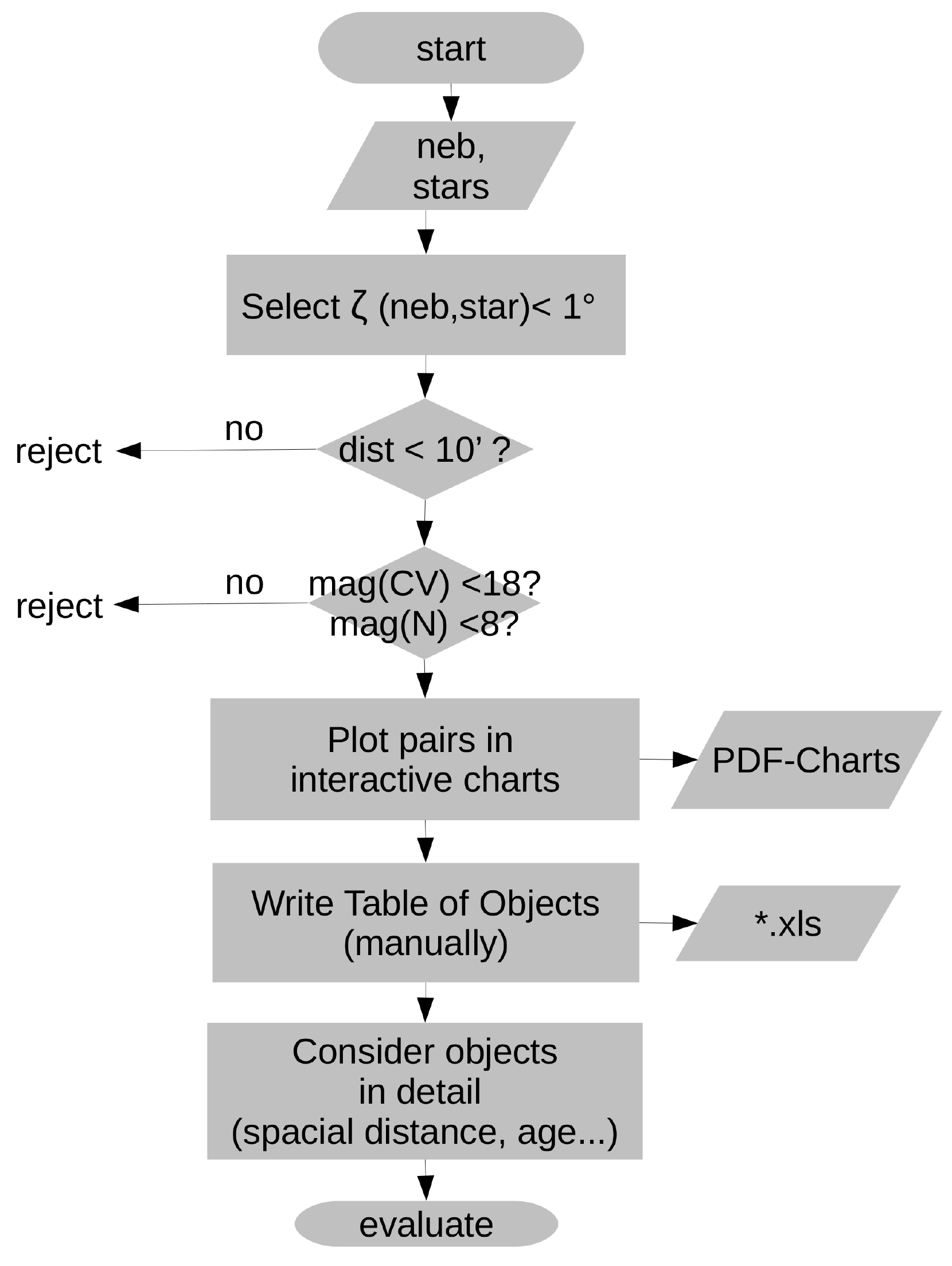} \\}
\end{figure}
 
 The algorithm of our analysis is displayed as flowchart in Fig.~\ref{fig:flow}. The apparent 2D distances are computed as angular separation $\zeta$ (orthodrome): 
 \begin{eqnarray} 
  \zeta =& \arccos\bigl( \sin DE_\textrm{neb} \sin DE_\textrm{star} + \nonumber \\
		& +\cos DE_\textrm{neb} \cos DE_\textrm{star} \cdot \cos(RA_\textrm{star} - RA_\textrm{neb}) \bigr), 
		\label{eq:dist}
 \label{equ:orthodrome} 
 \end{eqnarray} 
where $RA$ is the right ascension, $DE$ the declination and the indices `neb' and `star' stand for the nebula classified as (possible) PN and the common envelope binary, respectively, and the star could be a CV, an X-ray binary or a symbiotic system.

 \textbf{Step 4:} After computing the angular separations of all pairs which are smaller than 1\degr , we threw out all binaries which are too faint to brighten up to naked eye visibility. For the CVs, we applied our limit of 18~mag defined in \citet{hoffmannVogtProtte,hovoMNRAS2020} while for the symbiotic stars, we argued above that to our current knowledge even $>14$~mag could unlikely become visible for naked eye observers. 

 \textbf{Step 5:} Finally, a careful individual revision of the objects in our search fields (Tabs.~\ref{tab:pn} and \ref{tab:snr+psr}) is needed. Therefore, we developed the following procedure: 
  \begin{enumerate} 
	\item Check the ATNF Pulsar Catalog \citep{atnf} for the characteristic age $\tau_c$ of the pulsar (which provides an upper limit for the time of the supernova), 
	\item Check the supernova remnant catalogue provided by the University of Manitoba \citep{manitoba} whether an age estimates of the SNRs in the field are given, 
	\item Find out or compute the kinematic age object classified as `PN': 
	 \begin{enumerate} 
		\item Check if the real expansion rate of the particular object has already been determined. 
		\item If yes: Find discussions in the literature (whether it is almost certainly a PN and not a nova shell, whether it could have experienced a DIN or rebirth). 
		\item If not: estimate age with typical rates:
			\subitem age estimate from typical expansion rates of planetary nebulae: 42~km\,s$^{-1}$, \citep{jacob2013} 
			\subitem age estimate in case of typical expansion rates of nova shells: $500$ to $1200$~km\,s$^{-1}$, \citep{valle1997} and 		
	\end{enumerate} 
	\item Check if this age could possibly fit the time of the historical record.  
 \end{enumerate} 

	\textbf{Step 6:} Conclude whether or not one of the PNe, PSRs or SNRs found in the field could fit better than the others or even better than the CVs found there and suggested in our \citep{hoffmannVogtProtte,hovoMNRAS2020}.  

 This conclusion is briefly summarised in Tab.~\ref{tab:results}.
 
 \subsection{Test of the Close-Pair-Method}
 The Tab.~\ref{tab:closest} displays the findings of apparent close pairs (separated $\leq50\arcsec$) all over the sphere: 14 out of 18 CVs are classified as novae, line~13 of this table is in our search field for 1437 and will be discussed later as chance coincidence. Also, the first five entries are all within the Andromeda Galaxy (M31). The case of GK~Per, the bright Nova~1901, shows that this object and its nova shell which is obviously misclassified as PN is found with an angular separation of 0\farcs03.   
 \begin{table*}
 \caption{Closest pairs of nebulae and CVs, i.\,e. pairs which are separated by less than 50\arcsec .}
 \label{tab:closest}
 \scriptsize
 \begin{tabular}{rlll lccl c}
 id & PN name &RA2000 &DE2000 & CV name &  CV type & const. & dist/ arcsec & commentary\\ \hline
 1 & \text{[MMD2006] 1458    } & 10.4142 & 41.0637 & \text{Rosino N 39} & \text{NR:} & \text{And} & 21.3567 & in M31\\
 2 & \text{[AMB2011] 140     } & 10.4208 & 41.0649 & \text{Rosino N 39} & \text{NR:} & \text{And} & 17.0276  & in M31\\
 3 & \text{[AMB2011] 148     } & 10.4388 & 41.059 & \text{Rosino N 39} & \text{NR:} & \text{And} & 49.7265  & in M31\\
 4 & \text{[MMD2006] 1441    } & 10.4229 & 41.0468 & \text{Rosino N 39} & \text{NR:} & \text{And} & 48.7448  & in M31\\
 5 & \text{[MMD2006] 1449    } & 10.4113 & 41.0577 & \text{Rosino N 39} & \text{NR:} & \text{And} & 26.6695 & in M31\\
 6 & \text{V* FT Cam         } & 50.3097 & 61.0907 & \text{FT Cam} & \text{UG} & \text{Cam} & 0.0335289 &\\
 7 & \text{V* GK Per         } & 52.8 & 43.9043 & \text{GK Per} & \text{NA/DQ+UG} & \text{Per} & 0.0328169 & Nova 1901\\
 8 & \text{EGB 4             } & 97.3915 & 71.0768 & \text{BZ Cam} & \text{NL/VY} & \text{Cam} & 0.877421 &\\
 9 & \text{V* CP Pup         } & 122.942 & -35.3514 & \text{CP Pup} & \text{NA/DQ} & \text{Pup} & 0.0449627 &\\
 10 & \text{PN G326.9+08.2   } & 229.576 & -47.6411 & \text{USNO-B1.0 0423-0551514} & \text{N:} & \text{Lup} & 0.677654 &\\
 11 & \text{PN HaTr 5        } & 255.367 & -43.0986 & \text{GDS$\_$J1701281-430612} & \text{NA+UG/DQ+E} & \text{Sco} & 17.2775 &\\
 12 & \text{NOVA Sco 1949    } & 261.535 & -39.0674 & \text{V0902 Sco} & \text{N:} & \text{Sco} & 2.12569 &known nova\\
 13 & \text{PN G009.8-07.5   } & 279.095 & -23.9217 & \text{OGLE-BLG-DN-1057} & \text{UG} & \text{Sgr} & 49.7593  & in M22 \\
 14 & \text{V* V4368 Sgr     } & 283.668 & -19.6999 & \text{V4368 Sgr} & \text{NC:} & \text{Sgr} & 0.114261 &\\
 15 & \text{V* HM Sge        } & 295.488 & 16.7444 & \text{HM Sge} & \text{NC+M} & \text{Sge} & 0.270039 & \\
 16 & \text{EM* AS 373       } & 299.271 & 39.8267 & \text{V1016 Cyg} & \text{NC+M} & \text{Cyg} & 0.329677 &\\
 17 & \text{V* V1329 Cyg     } & 312.755 & 35.5817 & \text{V1329 Cyg} & \text{NC+E} & \text{Cyg} & 1.02141 &\\
 18 & \text{PN G088.0+00.4   } & 315.52 & 47.1708 & \text{IPHAS J210205.83+471018.0} & \text{NL} & \text{Cyg} & 11.8268 &\\
 \end{tabular} 
 \end{table*}

 We resume, that the method of searching close pairs of nebulae and CVs generally works fine but returns several misclassification and chance coincidences in areas with high apparent object density. Concluding, it requires a careful handpicking after the computation which provides a first filter. This discussion is presented in Section~\ref{chap:disc}. 
 
 \subsection{Distribution of our targets and expectations}
\begin{figure}
    \caption{Distribution of supernova remnants and pulsars (top) and nebulae classified as (possible or true) planetary nebulae (bottom) in Simbad, equatorial coordinates, equinox~2000. Clearly visible is a concentration of the nebulae in low Galactic latitudes. SNR: red `O', PSR: blue $\ast$, PN: cyan $\odot$, possible PN: light blue $\otimes$.}
    \label{fig:distrSNR+PN}
	\includegraphics[width=\columnwidth]{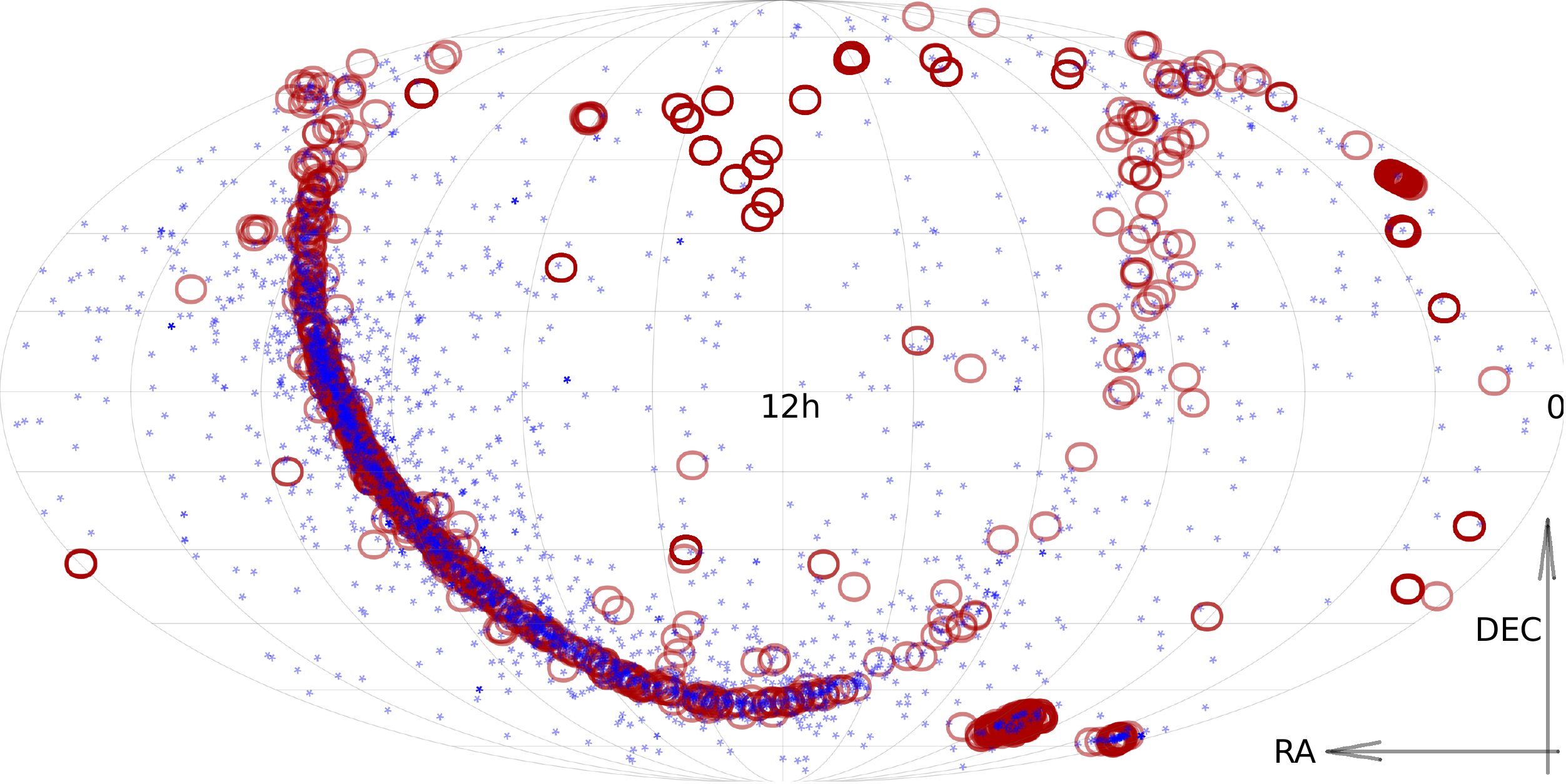}\\ 
	\includegraphics[width=\columnwidth]{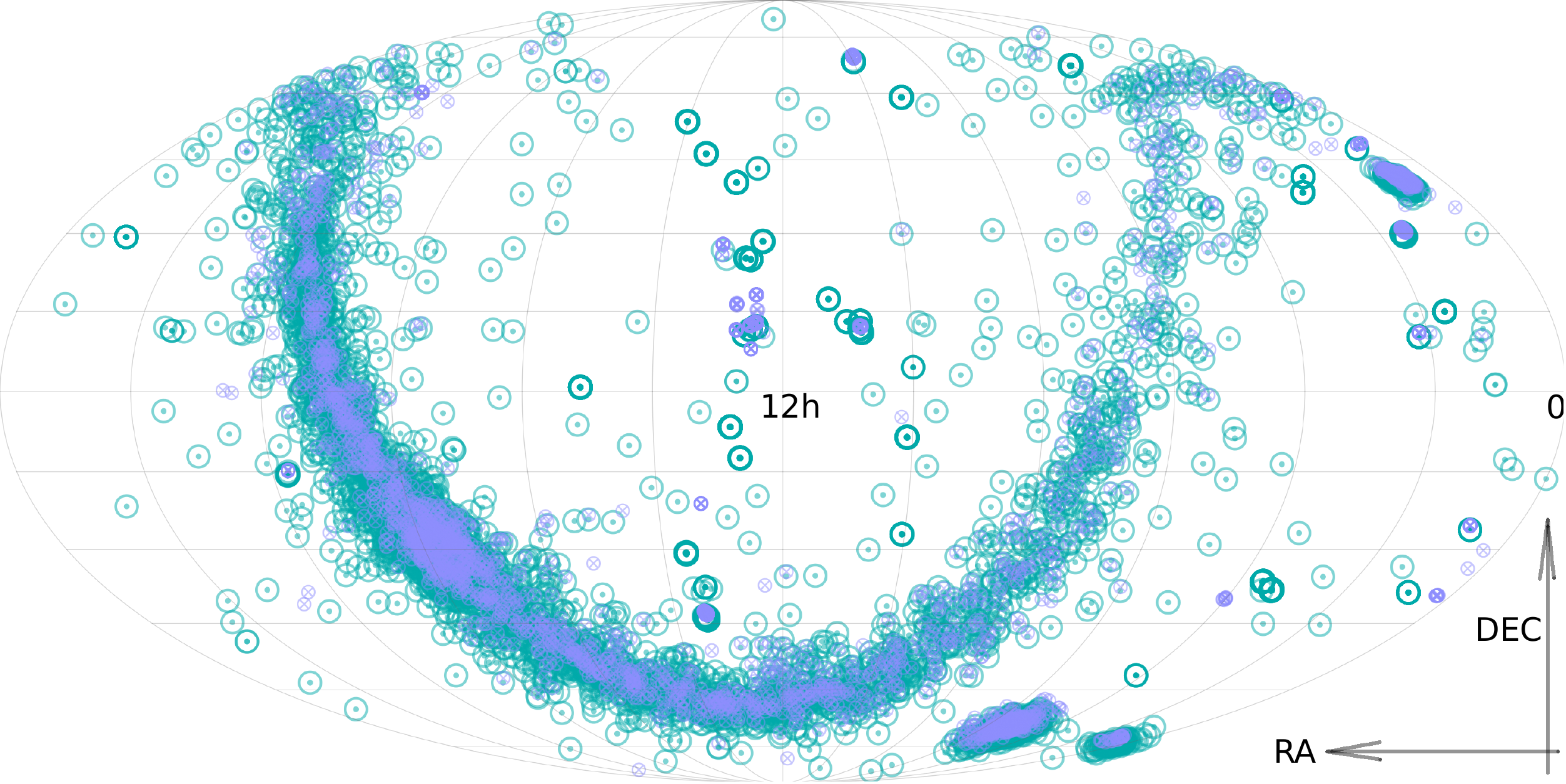}	
\end{figure}

 The cataclysmic variables are distributed almost isotropically on the celestial sphere. In our previous papers, therefore, our statistical expectations from \citet{vogt2019} matched the totals of CVs in our search fields in  \citet{hovoMNRAS2020}. With regard to the given distribution, this will not be the case for the consideration of PNe and SNRs because of their stronger concentration at low Galactic latitudes: see Fig.~\ref{fig:distrSNR+PN}. 

 Fig.~\ref{fig:distrSNR+PN} shows the distribution of known SNRs and PSRs on the sphere: While more SNRs are known in low Galactic latitudes, PSRs are expected to distribute isotropically because within their lifetime of usually $10^6 - 10^7$~years they can travel far away from the Galactic plane. Our knowledge might only be biased by observational methods such as the range and field of view of our surveying instruments, cf. e.\,g. \citet{hulse1974,nice1995,burgay2006,levin2013} on Arecibo observations. 

\begin{figure}
    \caption{The distribution of the considered planetary ($\odot$) nebulae with nearby CV ($\diamondsuit$) and our search fields (circles) in rectangular projection. The concentration of PN in vicinity of the Galactic plane and the broader distribution of our search fields explains the small profit of this search compared to the number of pairs all over the sphere (equinox:\,2000).}
    \label{fig:distPairs}
	\includegraphics[width=\columnwidth]{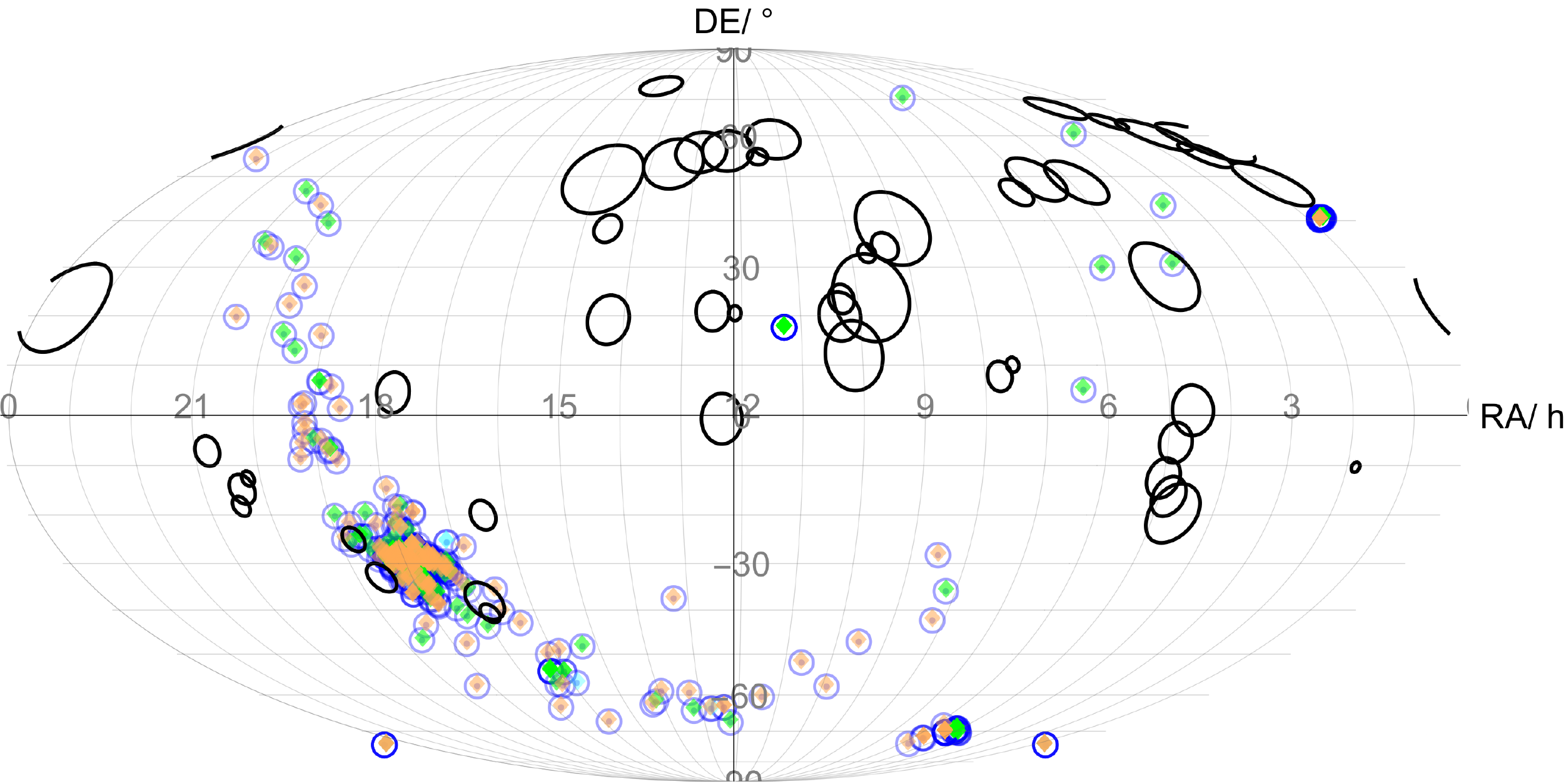} 
\end{figure}

 In our search fields (cf. Fig.~\ref{fig:distPairs}), we expect to find only few SNRs but some PSRs.  

 This distribution of PN and PN candidates also effects our findings of pairs of small angular separations: Of course, they will have a higher density at the low Galactic latitudes where more nebulae of any type are known. 
  
 On the whole sphere, we found a total of 42,006 pairs of PN-CV with angular separations $<1\degr$ and 2,561 PN-symbiotic pairs. This number is higher than the total of symbiotic stars (286) because at dense areas such as the Galactic bulge, M\,31 and the Magellanic clouds, some stars are close to more than one PN centre. Among the $42,006+2,561$ close pairs, $5,617+233$ pairs have distances smaller than 10\arcmin. The map in Fig.~\ref{fig:distPairs} shows their distribution on the sphere and with respect to our search fields. 

 The comparison of the places of our search fields with the distribution of close pairs (Fig~\ref{fig:distPairs}) and with the distributions of the nebulae (Fig.~\ref{fig:distrSNR+PN}) explains the brevity of object lists for the considered historical events. 
 
 \section{Resulting datasets for the historical analysis}
  The lists of PNe, SNRs, and PSRs in our defined search fields for the historical events was generated manually in the following way: We plotted our star charts for these fields by using our maps (cf. Appendices) in interactive Computable Document Format (CDF) displaying the names of the objects within our search fields with mouse-over labels. 

Then, we denoted the names of PNe and PN candidates in the search fields (manually) and added to this list the data from the original list from the Simbad query in order to obtain our selection table. The resulting lists of PN in our search fields is given in Table~\ref{tab:pn}. This process of plotting all known nebulae into the charts with our search circles also allows to see and judge quickly if the object in the circle really fulfills further conditions where applicable.\footnote{In \citet{hoffmannVogtProtte} on the CV search, we needed to check this in an additional step of evaluation after the catalogue query.} Additionally, it allows to see the density of this object type in the considered area and to see if there is an object slightly outside our search circle which could possibly also be taken into account -- by turning a blind eye to the limits of our intended search fields. 
\begin{table*}
	\centering
	\begin{minipage}{\textwidth}
	\caption{Complete list of Planetary Nebulae and (uncertain) candidates for PN in the search fields. These objects are discussed in Sect.~\ref{chap:disc}.}
	\label{tab:pn}
	\scriptsize
	\begin{tabular}{rrrlccrrp{5.5cm}} 
		\hline
		id & \text{year} & id$_y$ & \text{name} & \text{type} & \text{Vmag} & \text{RA2000} & \text{DE2000} & \text{commentary} \\
		\hline
		 1 & -47. & 1. & \text{PN M 3-32                      } & \text{PN } & \text{     $\sim $} & 281.17929 & -25.35967 & \text{} \\
 2 & -47. & 2. & \text{PN M 3-33                      } & \text{PN } & 13.2 & 282.0505856040811 & -25.4812284601227 & \text{} \\
 3 & -47. & 3. & \text{PN H 2-48                      } & \text{PN } & 13. & 281.6463291650335 & -23.4467352040472 & \text{} \\
 4 & -47. & 4. & \text{PTB 32                         } & \text{PN } & \text{     $\sim $} & 281.29250 & -23.36103 & \text{} \\
 5 & -47. & 5. & \text{EM* AS 327                     } & \text{Sy*} & 13. & 283.3194315721339 & -24.3830013641234 & \text{not a PN} \\
 6 & 70. & 1. & \text{IRC +10216                     } & \text{C* } & 10.96 & 146.989192 & +13.278767 & \text{} \\
 7 & 70. & 2. & \text{EGB 6                          } & \text{PN } & 15.999 & 148.2457782097270 & +13.7429512703335 & \text{} \\
 8 & 70. & 3. & \text{[MCG2003] Leo A PN 1           } & \text{PN?} & \text{     $\sim $} & 149.882 & 30.7578 & \text{} \\
 9 & 70. & 4. & \text{[SHK91] Leo A 3                } & \text{PN } & \text{     $\sim $} & 150.1292 & +30.7567 & \text{} \\
 10 & 70. & 5. & \text{[MCW2002] SexB PN1-4            } & \text{PN } & \text{     $\sim $} & 149.98542 & +05.32486 & \text{} \\
 11 & 329. & 1. & \text{M 97                           } & \text{PN } & 15.777 & 168.6988020921600 & +55.0190237014077 & \text{fits well the text position} \\
 12 & 641. & 1. & \text{[JCF90] NGC 4382 1-102           } & \text{PN?} & \text{     $\sim $} & 186.364 & 18.1853 & \text{NGC 4382 is a galaxy with plenty of PN} \\
 13 & 667. & 1. & \text{V* V1023 Tau                   } & \text{TT*} & 12. & 064.6959567123343 & +28.3354139798600 & \text{likely not PN (but T Tau)} \\
 14 & 667. & 2. & \text{NGC 1514                       } & \text{PN } & 9.48 & 062.3207738739905 & +30.7759638713768 & \text{CSPN} \\
 15 & 667. & 3. & \text{HD 281789                      } & \text{*  } & 10.16 & 063.3334338780734 & +31.1798011515431 & \text{not PN but type A1 star} \\
 16 & 667. & 4. & \text{PN K 3-66                      } & \text{PN } & \text{     $\sim $} & 069.15512 & +33.65833 & \text{pre-PN with dust lane} \\
 17 & 667. & 5. & \text{PN H 3-29                      } & \text{PN } & \text{     $\sim $} & 069.348188 & +25.045528 & \text{PN with dust} \\
 18 & 667. & 6. & \text{CoKu Tau-Aur Star 4            } & \text{TT*} & 14.68 & 070.3200442605474 & +28.6666881037394 & \text{likely not PN (but T Tau)} \\
 19 & 667. & 7. & \text{2MASS J04595608+2706020        } & \text{Sy2} & \text{     $\sim $} & 074.983696 & +27.100594 & \text{Seyfert galaxy} \\
 20 & 668. & 1. & \text{NGC 2242                       } & \text{PN } & \text{     $\sim $} & 98.5307 & 44.7772 & \text{} \\
 21 & 683. & 1. & \text{IC 2149                        } & \text{PN } & 11.35 & 89.0996 & 46.1048 & \text{too old} $4\cdot10^5$~years \\
 22 & 722. & 1. & \text{PN K 3-93                      } & \text{PN } & \text{     $\sim $} & 036.625063 & +65.798081 & \text{} \\
 23 & 722. & 2. & \text{PN K 3-91                      } & \text{PN } & \text{     $\sim $} & 029.650725 & +66.566147 & \text{} \\
 24 & 722. & 3. & \text{PN K 3-92                      } & \text{PN } & \text{     $\sim $} & 030.921546 & +64.960522 & \text{} \\
 25 & 722. & 4. & \text{NAME PN Fe 6                   } & \text{PN } & \text{     $\sim $} & 029.1038 & +65.4750 & \text{} \\
 26 & 722. & 5. & \text{PN A66 30                      } & \text{PN } & 14.3 & 131.7227668268712 & +17.8795466720606 & \text{} \\
 27 & 722. & 6. & \text{IC 1747                        } & \text{PN } & 15.4 & 029.3987080597163 & +63.3217898905825 & \text{} \\
 28 & 722. & 7. & \text{IRAS 01224+6120                } & \text{PN } & \text{     $\sim $} & 021.4359288284827 & +61.6032422673801 & \text{} \\
 29 & 722. & 8. & \text{PN G127.6-01.1                 } & \text{PN } & \text{     $\sim $} & 022.7871 & +61.3828 & \text{} \\
 30 & 722. & 9. & \text{PN K 4-59                      } & \text{Em*} & \text{     $\sim $} & 022.6378227621071 & +60.5220821945185 & \text{} \\
 31 & 722. & 10. & \text{PN G129.2-02.0                 } & \text{PN } & \text{     $\sim $} & 025.658333 & +60.162367 & \text{} \\
 32 & 722. & 11. & \text{SH 2-188                       } & \text{PN } & 17.447 & 022.6382406015929 & +58.4139790323242 & \text{} \\
 33 & 722. & 12. & \text{PN WeSb 1                      } & \text{PN } & \text{     $\sim $} & 015.222129 & +55.063303 & \text{} \\
 34 & 722. & 13. & \text{[CMG2005] NGC 185 4            } & \text{PN } & \text{     $\sim $} & 009.7348586241542 & +48.3225862322030 & \text{} \\
 35 & 722. & 14. & \text{[CMG2005] NGC 147 1-8            } & \text{PN?} & \text{     $\sim $} & 8.26575 & 48.5237 & \text{} \\
 36 & 722. & 15. & \text{SDSS J005232.04+440613.5       } & \text{PN?} & \text{     $\sim $} & 13.1335 & 44.1038 & \text{} \\
 37 & 722. & 16. & \text{[MMD2006] 2429                 } & \text{PN?} & 23.08 & 11.7817 & 43.0729 & \text{} \\
 38 & 722. & 17. & \text{[MMD2006] 2435                 } & \text{PN?} & 24.72 & 12.6429 & 42.9483 & \text{} \\
 39 & 722. & 18. & \text{NAME PN M31-372                } & \text{PN } & \text{     $\sim $} & 11.6729 & 43.9844 & \text{} \\
 40 & 722. & 19. & \text{SDSS J005247.50+442257.7       } & \text{PN } & \text{     $\sim $} & 13.1979 & 44.3827 & \text{} \\
 41 & 722. & 20. & \text{NAME PN Fe 8                   } & \text{PN?} & \text{     $\sim $} & 32.5443 & 65.4209 & \text{} \\
 42 & 722. & 21. & \text{IRAS 01571+6018                } & \text{*  } & \text{     $\sim $} & 30.1645 & 60.5498 & \text{} \\
 43 & 722. & 22. & \text{2MASS J01455120+6416057        } & \text{PN?} & \text{     $\sim $} & 26.4633 & 64.2682 & \text{} \\
 44 & 722. & 23. & \text{NAME PN Ra 18                  } & \text{PN?} & \text{     $\sim $} & 28.9125 & 62.6722 & \text{} \\
 45 & 722. & 24. & \text{NAME PN Mul 8                  } & \text{PN?} & \text{     $\sim $} & 24.6791 & 55.0829 & \text{} \\
 46 & 840. & 1. & \text{RAFGL 3068                     } & \text{C* } & \text{     $\sim $} & 349.802529 & +17.192536 & \text{LL Peg=Carbon Star of Mira type in PN} \\
 47 & 840. & 2. & \text{PN Jn 1                        } & \text{PN } & 15.62 & 353.9721733351536 & +30.4684270240498 & old PN \citep{napiwotzki1995} \\
 48 & 1175. & 1. & \text{PG 1520+525                    } & \text{PN } & 16.6 & 230.4440158845437 & +52.3677413094009 & \text{too old} $4\cdot10^4$~years \\
 49 & 1430. & 1. & \text{PN HDW 7                       } & \text{PN } & 17.6 & 118.7971141181319 & +09.5525499483599 & \text{interesting} \\
 50 & 1431. & 1. & \text{PN A66 73                      } & \text{PN } & \text{     $\sim $} & 314.112633 & +57.434181 & \text{} \\
 51 & 1431. & 2. & \text{NGC 1909                       } & \text{RNe} & \text{     $\sim $} & 075.5000 & -07.9000 & \text{Witch Head:} too big \\
 52 & 1431. & 3. & \text{PN MaC 2-1                     } & \text{PN } & 14.4 & 075.9243861695177 & -06.1675081203768 & \text{} \\
 53 & 1437. & 1. & \text{PN PM 1-114                    } & \text{PN?} & \text{     $\sim $} & 252.8252129958691 & -41.2210968481807 & \text{fits well the text position, not certain PN } \\
 54 & 1437. & 2. & \text{PN G344.4+01.8                 } & \text{PN?} & \text{     $\sim $} & 253.68037 & -40.69636 & \text{fits well the text position, in MASH II, not certain PN } \\
 55 & 1437. & 3. & \text{PN G344.0+02.5                 } & \text{PN } & \text{     $\sim $} & 252.58425 & -40.50086 & \text{fits well the text position, in MASH II} \\
 56 & 1437. & 4. & \text{ESO 332-4                      } & \text{PN } & \text{     $\sim $} & 252.889896 & -40.048892 & \text{} \\
 57 & 1437. & 5. & \text{BMP J1651-3930                 } & \text{PN } & \text{     $\sim $} & 252.9221153777408 & -39.5076339705931 & \text{} \\
 58 & 1437. & 6. & \text{ESO 332-2                      } & \text{PN } & \text{     $\sim $} & 252.38687 & -39.35253 & \text{} \\
 59 & 1437. & 7. & \text{PN G344.2+01.6                 } & \text{PN } & \text{     $\sim $} & 253.75208 & -40.92661 & \text{fits well the text position, almost nothing known} \\
 60 & 1437. & 8. & \text{PN Vd 4                        } & \text{PN } & 15.5 & 252.6056215316712 & -39.1386394581621 & \text{} \\
 61 & 1437. & 9. & \text{PN G345.9+03.0 a               } & \text{PN } & \text{     $\sim $} & 253.6358 & -38.7508 & \text{} \\
 62 & 1437. & 10. & \text{PN G345.8+02.7                 } & \text{PN } & \text{     $\sim $} & 253.9666794759882 & -39.0059781135805 & \text{} \\
 63 & 1437. & 11. & \text{PN G345.8+02.4                 } & \text{PN } & \text{     $\sim $} & 254.167158 & -39.210269 & \text{} \\
 64 & 1437. & 12. & \text{PN G346.1+02.8                 } & \text{PN?} & \text{     $\sim $} & 254.0071 & -38.6975 & \text{} \\
 65 & 1437. & 13. & \text{PN G346.5+02.7                 } & \text{PN } & \text{     $\sim $} & 254.4367 & -38.4464 & \text{} \\
 66 & 1437. & 14. & \text{PN G343.5+01.2                 } & \text{PN } & \text{     $\sim $} & 253.4804 & -41.7333 & \text{} \\
 67 & 1437. & 15. & \text{PN G343.6+01.1                 } & \text{PN?} & \text{     $\sim $} & 253.7125 & -41.7306 & \text{} \\
 68 & 1437. & 16. & \text{PN H 1-5                       } & \text{PN } & 15. & 254.3489570106828 & -41.6328327176266 & \text{} \\
 69 & 1437. & 17. & \text{PN G343.6+03.7 a               } & \text{PN } & \text{     $\sim $} & 251.0850 & -40.0369 & \text{} \\
 70 & 1437. & 18. & \text{IRAS 16456-3822                } & \text{PN } & \text{     $\sim $} & 252.2567262544107 & -38.4620085992982 & \text{} \\
 71 & 1437. & 19. & \text{ESO 332-18                     } & \text{PN } & \text{     $\sim $} & 256.1408091875673 & -37.8874542478312 & \text{} \\
 72 & 1461.	&	 & \text{PN DeHt 4}                  				&PN & \text{     $\sim $}  &291.611100	&+13.326308 & \\
 73 & 1690. & 1. & \text{IC 4776                        } & \text{PN } & 10.62 & 281.461271 & -33.342811 & \text{post-CE nebula! interesting case} \\
 74 & 1690. & 2. & \text{LSE 63                         } & \text{pA*} & 11.97 & 280.0917439775203 & -31.9468926340524 & \text{Post-AGB Star (proto-PN)} \\
 75 & 1690. & 3. & \text{PN G002.8-10.7                 } & \text{PN } & \text{     $\sim $} & 278.8900 & -31.5958 & \text{true PN (MASH)} \\
 76 & 1690. & 4. & \text{PHR J1833-3115                 } & \text{PN } & \text{     $\sim $} & 278.42838 & -31.26192 & \text{true PN (MASH)}, $225\arcsec\times35\arcsec$ \\
 77 & 1690. & 5. & \text{PN Cn 1-5                      } & \text{PN } & 15.2 & 277.2985433345579 & -31.4997569852074 & \text{CSPN WC} \\
 78 & 1690. & 6. & \text{PN G002.1-08.3                 } & \text{PN?} & \text{     $\sim $} & 276.0904 & -31.1194 & \text{likely PN, $9\arcsec\times7\arcsec$ (weak NII), Em, strong H}$\alpha$ \\
 79 & 1690. & 7. & \text{PN M 3-29                      } & \text{PN } & 14.1 & 279.8575767162189 & -30.6770067167098 & \text{0.137 0.137 arcm} \\
 80 & 1690. & 8. & \text{PN G001.1-11.5                 } & \text{PN?} & \text{     $\sim $} & 278.9308 & -33.4125 & \text{possible PN, $5\arcsec\times4\arcsec$, S, no continuum, weak H$\alpha$, spec noisy} \\
 81 & 1690. & 9. & \text{PN SB 55                       } & \text{PN } & \text{     $\sim $} & 274.859712 & -33.618017 & \text{Frew, 2013} \\

		\hline
	\end{tabular}
	\end{minipage} 
\end{table*}

 A similar algorithm we applied for supernova remnants and for pulsars. The complete dataset downloaded from CDS Simbad was plotted in maps of CDF format and together with our search circles. These maps are shown in Appendix~B and the resulting list of objects in these fields is displayed in Table~\ref{tab:snr+psr}.  
\begin{table*}
	\centering
	\begin{minipage}{\textwidth}
	\caption{Complete list of Supernova Remnants and pulsars in the search fields. These objects are discussed in Sect.~\ref{chap:disc}.}
	\label{tab:snr+psr}
	\scriptsize
	\begin{tabular} {rrrlrrccp{3.5cm}} 
		\hline
		 id  & \text{year} & id$_y$ & \text{name} & \text{RA2000} & \text{DE2000} & \text{age/yr} & \text{P/s} & \text{commentary} \\
		\hline
 1 & $-4$ & 1 & \text{PSR J2010-1323 } & 302.691 & $-13.3989$ & $1.72\times 10^{10}$ & \text{} & \text{too old} \\
 2 & 70 & 1 & \text{[MPW94] 43.74+62.5           } & 148.969 & 69.6802 & $\ast$ & 0.535829 & \text{} \\
 3 & 70 & 2 & \text{PSR J1022+10             } & 155.742 & 10.0313 & \text{not in ATNF} & \text{} & \text{} \\
 4 & 70 & 3 & \text{PSR J1000+08             } & 150.158 & 8.3328 & $\ast$ & 0.440372 & \text{old } \\
 5 & 70 & 4 & \text{PSR B0950+08             } & 148.289 & 7.9266 & $1.75\times 10^7$ & \text{} & \text{old } \\
 6 & 70 & 5 & \text{PSR B0943+10             } & 146.53 & 9.86592 & $4.98\times 10^6$ & \text{} & \text{old } \\
 7 & 70 & 6 & \text{PSR J1010+15             } & 152.5 & 15.85 & $\ast$ & $\ast$ & \text{} \\
 8 & 70 & 7 & \text{PSR B0940+16             } & 145.875 & 16.5269 & $1.89\times 10^8$ & \text{} & \text{too old} \\
 9 & 70 & 8 & \text{PSR J0943+22             } & 145.854 & 22.9333 & \text{not in ATNF} & 0.532913 & \text{Ray+,ApJ1996} \\
 10 & 70 & 9 & \text{PSR J0927+23             } & 141.904 & 23.7833 & \text{not in  ATNF} & 0.761886 & \text{Ray+,ApJ1996} \\
 11 & 70 & 10 & \text{PSR J0947+27             } & 146.842 & 27.7 & \text{not in  ATNF} & 0.85105 & \text{Ray+,ApJ1996} \\
 12 & 70 & 11 & \text{PSR J1005+3015           } & 151.375 & 30.25 & \text{not in ATNF} & \text{} & \text{} \\
 13 & 70 & 12 & \text{PSR J0928+3037           } & 142.179 & 30.6167 & \text{not in ATNF} & \text{} & \text{} \\
 14 & 70 & 13 & \text{PSR J0926+3018           } & 141.5 & 30.3 & \text{not in ATNF} & \text{} & \text{} \\
 15 & 70 & 14 & \text{PSR J0943+4109           } & 145.75 & 41.15 & \text{not in ATNF} & \text{} & \text{Thorsett+,ApJ,1993} \\
 16 & 70 & 15 & \text{PSR J0933+3245           } & 143.458 & 32.75 & \text{not in ATNF} & \text{} & \text{} \\
 17 & 641 & 1 & \text{PSR J1238+21             } & 189.597 & 21.8698 &  $\sim10^7$ & 1.11836 & \text{Ray+,ApJ1996, Hobbs+,MNRAS,2004} \\
 18 & 667 & 1 & \text{PSR J0457+23             } & 74.275 & 23.5667 & $\ast$ & 0.5049 & \text{old } \\
 19 & 667 & 2 & \text{PSR J0435+27             } & 68.8917 & 27.7333 & \text{not in ATNF} & 0.326279 & \text{Ray+,ApJ1996} \\
 20 & 667 & 3 & \text{PSR J0421+3240           } & 65.375 & 32.6667 & \text{not in ATNF} & \text{} & \text{} \\
 21 & 668 & 1 & \text{2XMM J050106.5+451634    } & 75.2833 & 45.2753 & \text{not in ATNF} & \text{} & \text{U Manitoba: SGR 0501+4516, magnetar, possibly related to SNR G160.9+02.6} \\
 22 & 668 & 1 & \text{SNR G160.1-01.1              } & \text{} & \text{} & \text{no context data} & \text{} & \text{} \\
 23 & 668 & 2 & \text{PSR J0426+4933           } & 66.5284 & 49.5607 & 371000 & \text{} & \text{} \\
 24 & 668 & 2 & \text{SNR G156.4-01.2              } & \text{} & \text{} & \text{not in U Manitoba} & \text{} & \text{} \\
 25 & 668 & 3 & \text{PSR B0458+46             } & 75.519 & 46.9017 &$ 1.81\times 10^6$ & \text{} & \text{old } \\
 26 & 668 & 3 & \text{SNR G160.8+02.6 =SNR G160.9+02.6} & \text{} & \text{} & \text{4000-7000} & \text{} & \text{PSR age 16000} \\
 27 & 668 & 4 & \text{SNR G156.2+05.7              } & \text{} & \text{} & 7000 $--$ 26000 & \text{} & \text{} \\
 28 & 668 & 5 & \text{SNR G159.6+07.3              } & \text{} & \text{} & \text{no data} & \text{} & \text{} \\
 29 & 683 & 1 & \text{SNR G159.6+07.3              } & 80 & 50 & \text{no data} & \text{} & \text{faint H$\alpha$, but $3\degr\times4\degr$ huge, not result of an event in 683} \\
 30 & 722 & 1 & \text{[GHM84] SNR                  } & 9.7397 & 48.3378 & \text{no data} & \text{} & \text{} \\
 31 & 722 & 1 & \text{PSR J0205+6449           } & 31.408 & 64.8286 & 5370 & \text{} & \text{possibly SN 1181} \\
 32 & 722 & 2 & \text{PSR B0154+61             } & 29.458 & 62.2072 & 197000 & \text{} & \text{} \\
 33 & 722 & 2 & \text{SNR G128.5+02.6              } & 26.0676 & 64.899 & \text{no data} & \text{} & \text{} \\
 34 & 722 & 3 & \text{PSR J0137+6349           } & 24.25 & 63.8167 & $1.19\times 10^7$ & \text{} & \text{old } \\
 35 & 722 & 3 & \text{SNR G127.3+00.7 $=$G127.1+00.5          } & 22.2 & 63.0667 & 20000--30000 & \text{} & \text{too old} \\
 36 & 722 & 4 & \text{EM$^\ast$ GGA 104              } & 26.7509 & 61.3566 & \text{not in ATNF} & \text{} & \text{HMXB with nebula (SDSS), interesting target!} \\
 37 & 722 & 4 & \text{SNR G126.2+01.6              } & 20.5 & 64.25 & 270000 & \text{} & \text{too old} \\
 38 & 722 & 5 & \text{PSR J0146+6145           } & 26.5925 & 61.7511 & 69100 & \text{} & \text{} \\
 39 & 722 & 5 & \text{SNR G130.7+03.1= SNR 3C 58  } & 31.4043 & 64.8283  & \text{} & \text{} & \text{possible remnant of SN 1181} \\
 40 & 722 & 6 & \text{PSR B0144+59             } & 26.9361 & 59.3676 & $1.21\times 10^7$ & \text{} & \text{old } \\
 41 & 722 & 7 & \text{PSR J2238+59             } & 339.561 & 59.08 & \text{not in ATNF} & \text{} & \text{} \\
 42 & 722 & 8 & \text{PSR B0136+57             } & 24.8323 & 58.2421 & 403000 & \text{} & \text{} \\
 43 & 722 & 9 & \text{PSR B0114+58             } & 19.4111 & 59.244 & 275000 & \text{} & \text{} \\
 44 & 722 & 10 & \text{[KKL2015] J0103+54       } & 15.9042 & 54.0333 & $\ast$ & 0.354304 & \text{} \\
 45 & 722 & 11 & \text{PSR B0052+51             } & 13.9391 & 51.2903 & $3.51\times 10^6$ & \text{} & \text{old } \\
 46 & 722 & 12 & \text{PSR J0106+4855           } & 16.6043 & 48.9311 & $3.08\times 10^6$ & \text{} & \text{old } \\
 47 & 722 & 13 & \text{PSR J0103+48             } & 15.75 & 48 & \text{not in ATNF} & \text{} & \text{} \\
 48 & 722 & 14 & \text{PSR B0053+47             } & 14.1063 & 47.9363 & $2.25\times 10^6$ & \text{} & \text{old } \\
 49 & 722 & 15 & \text{PSR B0011+47             } & 3.57396 & 47.7759 & $3.48\times 10^7$ & \text{} & \text{old } \\
 50 & 840 & 1 & \text{PSR J2355+2246           } & 358.958 & 22.7714 & $7.72\times 10^6$ & \text{} & \text{old } \\
 51 & 840 & 2 & \text{PSR J2329+16             } & 352.458 & 16.95 & $\ast$ & 0.6321 & \text{old } \\
 52 & 840 & 3 & \text{PSR J2322+2057           } & 350.593 & 20.9508 & $7.89\times 10^9$ & \text{} & \text{too old} \\
 53 & 840 & 4 & \text{PSR J0006+1834           } & 1.52 & 18.5831 & $5.24\times 10^6$ & \text{} & \text{old } \\
 54 & 840 & 5 & \text{PSR J2317+1439           } & 349.288 & 14.6587 & $2.25\times 10^{10}$ & \text{} & \text{too old} \\
 55 & 840 & 6 & \text{PSR B2315+21             } & 349.491 & 21.83 & $2.19\times 10^7$ & \text{} & \text{old } \\
 56 & 840 & 7 & \text{PSR J2307+2225           } & 346.922 & 22.4306 & $9.76\times 10^8$ & \text{} & \text{too old} \\
 57 & 840 & 8 & \text{PSR J2317+29             } & 349.25 & 29 & \text{not in ATNF} & \text{} & \text{} \\
 58 & 891 & 1 & \text{PSR J1633-2010           } & 248.48 & $-20.1692$ & \text{not in ATNF} & \text{} & \text{} \\
 59 & 1175 & 1 & \text{PSR J1518+4904           } & 229.57 & 49.0762 & $2.39\times 10^{10}$ & \text{} & \text{too old} \\
 60 & 1175 & 2 & \text{PSR B1508+55             } & 227.357 & 55.5258 & $2.34\times 10^6$ & \text{} & \text{old } \\
 61 & 1175 & 3 & \text{PSR J1544+4937           } & 236.017 & 49.6326 & $1.17\times 10^{10}$ & \text{} & \text{too old} \\
 62 & 1431 & 1 & \text{PSR J0458-0505           } & 74.6547 & $-5.08475$ & $5.63\times 10^7$ & \text{} & \text{old } \\
 63 & 1431 & 2 & \text{PSR B0450-18             } & 73.1421 & $-17.9898$ & $1.51\times 10^6$ & \text{} & \text{old } \\
 64 & 1431 & 3 & \text{PSR B0447-12             } & 72.5366 & $-12.802$ & $6.76\times 10^7$ & \text{} & \text{old } \\
 65 & 1431 & 4 & \text{[KKL2015] J0447-04       } & 71.75 & $-4.5833$ & $\ast$ & $2.18819$ & \text{old } \\
 66 & 1431 & 5 & \text{PSR J0459-0210           } & 74.9664 & $-2.1685$ & $1.28\times 10^7$ & \text{} & \text{old } \\
 67 & 1437 & 1 & \text{PSR J1654-4140           } & 253.598 & $-41.6733$ & $1.55\times 10^8$ & \text{} & \text{fits text position very well but too old} \\
 68 & 1437 & 2 & \text{PSR J1650-4126           } & 252.555 & $-41.4432$ & $2.47\times 10^8$ & \text{} & \text{fits text position very well but too old} \\
 69 & 1437 & 3 & \text{PSR J1653-4030           } & 253.393 & $-40.5004$ & $3.7\times 10^7$ & \text{} & \text{fits text position very well} \\
 70 & 1437 & 4 & \text{PSR J1649-3935           } & 252.278 & $-39.5956$ & $3.12\times 10^8$ & \text{} & \text{too old} \\
 71 & 1437 & 5 & \text{PSR B1650-38             } & 253.416 & $-38.6391$ & $1.73\times 10^6$ & \text{} & \text{old } \\
 72 & 1437 & 6 & \text{PSR J1655-3844           } & 253.911 & $-38.7358$ & $9.45\times 10^6$ & \text{} & \text{old } \\
 73 & 1437 & 7 & \text{PSR J1700-4012           } & 255.161 & $-40.2107$ & $4.54\times 10^7$ & \text{} & \text{old } \\
 74 & 1437 & 8 & \text{PSR J1700-3919           } & 255.093 & $-39.3167$ & $1.77\times 10^9$ & \text{} & \text{too old} \\
	\end{tabular}
	\end{minipage} 
\end{table*} %
\begin{table*}
	\centering
	\begin{minipage}{\textwidth}
	\contcaption{ }
	\scriptsize
	\begin{tabular}{rrrlccrrp{3.5cm}} 
		\hline
		 id  & \text{year} & id$_y$ & \text{name} & \text{RA2000} & \text{DE2000} & \text{age/yr} & \text{P/s} & \text{commentary} \\
		\hline
 75 & 1497 & 1 & \text{PSR J1434+7257           } & 218.499 & 72.9574 & $1.2\times 10^9$ & \text{} & \text{too old} \\
 76 & 1497 & 2 & \text{[KKL2015] J1439+76       } & 219.75 & 76.9167 & $\ast$ & 0.947903 & \text{} \\
 77 & 1461 & 1 & \text{PSR J1738+04 }           &	264.6042	&4.3333	& $\ast$ 	&1.39179	&1.08 kpc \\
 78 & 1461 & 2 & \text{PSR J1743+05 }            &	265.8167	&5.4833	& $\ast$ 	&1.47363	 &3.95 kpc\\
 79 & 1461 & 3 & \text{PSR J1736+05 }            &	264.225		&5.8	& $\ast$ 	&0.999245	&2.49 kpc\\
 80 & 1461 & 4 & \text{PSR J1739+0612 }          &	264.824858	&6.207889	&$2.37\times 10^7$	&0.234169035616	&25 kpc\\
 81 & 1461 & 5 & \text{PSR J1750+07  }           &	267.6667	&7.55	& $\ast$ 	&1.90881	&3.75 kpc\\
 82 & 1461 & 6 & \text{PSR J1738+0333}           &	264.72484654	&3.55301083	&$3.84\times10^9$	&0.005850095859776	&MSP at 1.47 kpc \\
 83 & 1661 & 1 & \text{V$\ast$ LY Aqr }            & 312.781 & $-8.4605$ & \text{not in ATNF} & \text{} & \text{MSP in binary} \\
 84 & 1661 & 2 & \text{PSR B2043-04  }     & 311.501 & $-4.35722$ & $1.67\times 10^7$ & \text{} & \text{} \\
		\hline
	\end{tabular}
	\end{minipage} 
\end{table*}
 \subsection{Supernova remnants and pulsars}
 The characteristic ages $\tau_c$ of the PSRs listed in Table~\ref{tab:snr+psr} are $10^5$ to $10^{10}$ years, among which we excluded those with $\tau_c>10^7$ as millisecond pulsars, cf. Sect.~\ref{chap:snrMeth}. The others are discussed in more detail in Sect.~\ref{chap:disc}. The object with the lowest $\tau_c$ is SNR 3C~58, the remnant suggested to be identified with SN~1181 \citet{kothes2013}; we will discuss this alternative in Sect.~\ref{chap:disc}. 

 \begin{table*}
	\centering
	\caption{Symbiotic binaries in the search fields. To the VSX output data of coordinates, period, and magnitude range, we added the distance from Gaia parallaxes and computed the absolute magnitude M. `VF' means `very fast'.}
	\label{tab:symb}
	\small
	\begin{tabular} {rrlllclrcrp{2.5cm}} 
		\hline
id	&year	&star name	&RA2000		&DE2000		&dist./ kpc  &Type	&Period/ d	&mag	&M (V) &	commentary\\
		\hline
1	&$-47$	&AS 327		&$283.319$	&$-24.383$	&--		&ZAND		&823	&$12.6 - 13.5$ V	&--	&too faint\\
2	&$-47$	&V5569 Sgr	&$282.515$	&$-26.404$	&4.4	&EA+ZAND+BE	&515	&$9.8 - 12.1$ V		&$-3.4$	&possible but not perfectly fitting\\
3	&$-47$	&V2601 Sgr	&$279.509$	&$-22.698$	&$\geq4$ 	&ZAND	&850	&$14.0 - 15.3$ p	&$\leq +1.0$	&too faint, slightly outside circle\\
4	&$-4$	&StHA 176	&$305.676$	&$-21.132$	&$\geq7$	&ZAND	&120.9	&$11.9 - 12.8$ V	&$\leq-2.3$	&possible\\
5	&722	&V0832 Cas	&$26.910$	&$+60.699$	&5.4	&ZAND:		&--		&$11.66 - 13.5$ V	&$-2$	 	&VF nova, $\leq2$~mag: possible\\
6	&722	&AX Per		&$24.095$	&$+54.251$	&3.4 	&ZAND+E		&680.83	&$9.5 - 12.8$ V 	&$-1.1$ 	&VF nova, $\leq2$~mag: possible\\
7	&840	&V0379 Peg	&$358.469$ 	&$+23.156$	&0.11	&ZAND:		&--		&$13.9 * - 18.5$ V	&$8.7$	&too faint\\
8	&891	&NSV 20558	&$243.873$	&$-22$		&--		&ZAND		&--		&$15.6 - 15.9$ V	&--		&too faint\\
9	&1431	&StHA 32	&$69.440$	&$-1.31997$	&8.3	&ZAND		&626	&$12.2 - 12.9$ V	&$-2.4$	&`shiny bright' event, too faint\\
10	&1431	&KT Eri		&$71.976$	&$-10.179$	&--		&NA+ZAND:+E:&737	&$5.4 - 16.5:$ V	&--	& see Section 4.3 \\  
11	&1437	&Hen 2-173	&$249.103$	&$-39.862$	&4.1	&ZAND+EA	&911	&$13.4 - 14.5:$ V	&$0.3$	&too faint, position doesn't fit\\
12	&1661	&StHA 180	&$309.836$	&$-5.288$ 	&$\geq 7$ 	&ZAND+R	&1494	&$12.45 - 12.76$ V	&$\leq -1.8$	&event $\sim0.5$~mag bright, unlikely \\
13	&1690	&V3804 Sgr	&$275.3699$	&$-31.535$	&--		&ZAND		&426	&$10.5 - 13.4$ V	&--	&position unlikely\\
		\hline
	\end{tabular}
\end{table*}
  \subsection{Symbiotic stars}
  To our current knowledge, there are 286 Z~And-type stars at the whole sky and 105 of them fulfill the naked eye brightness conditions if they undergo today a typical nova eruption. Approximately half of them, i.\,e. $\sim50$ of the bright ones, have a nebula nearby. 
  
  The corresponding search in our search fields with the VSX-extracted dataset of symbiotic binaries led to 13 Z~And-type stars in our search fields out of which at least five are definitely too faint because they do not pass our 14~mag-limit for being able to brighten up to naked eye visibility. They are labeled as `too faint' in the last column of Tab.~\ref{tab:symb}. 

 All 13 binaries are listed together with their distances and estimates of absolute $V$ magnitude at maximum light. Distances are calculated from the parallaxes of the \citet{brown} for targets with reasonable accuracy (error $\sim10$\,\% of the parallax value). For targets with errors up to 30\,\% of their parallax values we only give minimal distances, based on the sum of the parallax value and its error. For the remaining targets in Tab.~\ref{tab:symb}, no useful Gaia parallaxes are available. 
All absolute magnitudes given in Tab.~\ref{tab:symb} refer to the average value of the  magnitude range listed in the VSX catalogue (see 9th column of this table); they should be considered as lower limits as no correction for the interstellar absorption was applied. 
 
 \subsection{Close apparent pairs of PN-CV in our search circles}
  Considering only the close pairs (distances $\leq10\arcmin$) and plotting them into our maps of the 22 most promising historical events, we found only seven pairs (see Tab.~\ref{tab:pairs}). The maps of the above mentioned pairs are depicted in Fig.~\ref{fig:pairs}. 
 \begin{table}
 \caption{Close pairs of nebulae and variable stars. Please find the maps of the according fields in Fig.~\ref{fig:pairs}. The $^\star$-symbol at the year indicates that this `couple' is obviously a match of a symbiotic star with itself.}
 \label{tab:pairs}
 \small
 \begin{tabular}{rlll}
 year &PN name &name of variable &   dist/ arcmin  \\\hline
 $-47^\star$	&EM$^\ast$ AS 327         &AS 327		 &0.0008  \\
$-47$		&PN M 3-33          &V0522 Sgr			 	&6.946\\
$-47$		&PN G009.8-07.5     &OGLE-BLG-DN-1057	 	&0.829 \\
$-47$		&PN G009.8-07.5     &OGLE-BLG-DN-1056	  	&4.599 \\
667		&HD 281789          &2MASS J04132921+3116279 &6.0131 \\
1437	&PN G349.7+04.0     &V1535 Sco		 		&9.469\\
1437$^\star$	&WRAY 15-1518       &Hen 2-173		 	&0.0001  \\
 \end{tabular} 
 \end{table}

\section{Discussion of the extracted data}
  In sum, we found the number of objects to be considered per event as displayed in Table~\ref{tab:findings}. 
 \begin{table}
	\centering
	\caption{Totals of SNRs, PSRs and PNe in the search fields per year as displayed in Tables~\ref{tab:pn} and \ref{tab:snr+psr}.}
	\label{tab:findings}
	\begin{tabular}{rrrrr} 
		\hline
		year	&PNe	 &close pairs &SNRs	&PSRs\\
		\hline
$-203$	&0	&0	&0 & 0\\
$-103$	&0	&0	&0 & 0\\
$-47$	&5 & 4	&0	&0\\
$-4$	&0 & 0	&0	&1 \\
64	&0	&0	&0 & 0\\
70	&5 & 0	&1	&14\\
101	&0 & 0	&0	&0\\
329	&1 & 0	&0	&0\\
641	&1 & 0	&0	&1\\
667	&7 & 1	&0	&3\\
668	&1 & 0	&5	&3\\
683	&1 & 0	&1	&0\\
722	&24 & 0	&5	&15\\
840	&2 & 0	&0	&8\\
891	&0 & 0	&0	&2\\
1175	&1 & 0	&0	&3\\
1430	&1 & 0	&0	&0\\
1431	&2(3) & 0	&0	&5\\
1437	&4 (51)  & 0 (2) &0	&3 (27)  \\
1461	&1 & 0  &0  &6\\
1497	&0 & 0	&0	&2\\
1661	&0 & 0	&0	&2\\
1690	&9 & 0	&0	&0\\
		\hline
	\end{tabular}
\end{table}
 These findings are discussed in the following subsections. Only one object of the above types is currently in the search fields of $-4$, $1430$, and $329$. Their likelihood in comparison with the CV findings of our earlier paper is discussed in the Section~\ref{chap:disc}.

 \subsection{Search fields with no findings}
For the appearances $-203$, $-103$, $+64$, and $101$, there are no currently known objects of any of the above mentioned types in the search fields. 

 Thus, for the \textbf{record of $+$101} we propose to accept the identification of the cataclysmic variables as we suggested in our earlier paper \citep{hovoMNRAS2020} with the still possible suggestion by \citet{hertzog1986,pat2013}. 

 \textbf{For $+$64} we suggested the cataclysmic binary V0379 Vir which appeared to be too faint to erupt up to naked eye visibility. Yet, it is still the only object found in this field and as it is too faint, we cannot offer any suggestion what might have caused this sighting.   

 The \textbf{appearance in $-$203} took place in the vicinity of the bright star Arcturus which is a single star asterism in Chinese sky culture. AB Boo is the brightest of three remaining CV candidates \citep{hovoMNRAS2020} according to our criteria specified in \citet{hoffmannVogtProtte} and was observed as Nova~1877. Although we did not find any PN or SNR or PSR within our search field, the PN SkAc~1 is only slightly outside our field; it is possible that this object contains a CV \citep{gentile2015}. This PN is located within a neighbouring asterism (see Fig.~A2). That means, the historical description `appeared in Dajioa' where Dajiao (the Great Horn) refers to Arcturus plus vicinity would only be correct if the astronomer did not use the full system of asterisms but only one of the (smaller and later merged) sets of fewer asterisms (possible in this epoch). However, the PN SkAc~1 has a much too big kinematic age (5,500 to 13,000 years in case of a nova shell, 160,000~years in case of a PN). Thus, we cannot provide any likely suggestion what could have caused this sighting. If AB~Boo as brightest candidate is recurrent within $\sim2000$~years, it would be worthwhile to look for a shell but the proximity of bright Arcturus makes this rather difficult. 

 The event given in the text has happened at a not specified time within the interval of three years `$-103$ to $-100$' and is reported at the `Zhaoyao' asterism which is commonly interpreted as the single star $\gamma$~Boo. As there is no known CV or X-ray binary in the search field \citep{hovoMNRAS2020} and also no PN, no known supernova remnant or pulsar, we possibly do not understand the brief historical note correctly: It is still possible that this `xing bo' (bushy star) of which no movement or tail is reported, was not a nova but a tailless comet (suggested as guest star only by \citet{xu2000} and neither by \citet{steph77} nor any other of the earlier authors). Alternatively, the interpretation of the star name is not correct. 

\subsection{Discussion of our `close pairs'}
\begin{figure}
    \caption{The three close PN-CV pairs within their search fields; further information: see text. The maps show our search circles and the Chinese asterism lines; at the upper rim the event year is given and in brackets the duration if given, at the right rim the position as described in the text is abbreviated. The coordinates are equatorial, equinox 2000.}
    \label{fig:pairs}
	\includegraphics[width=.88\columnwidth]{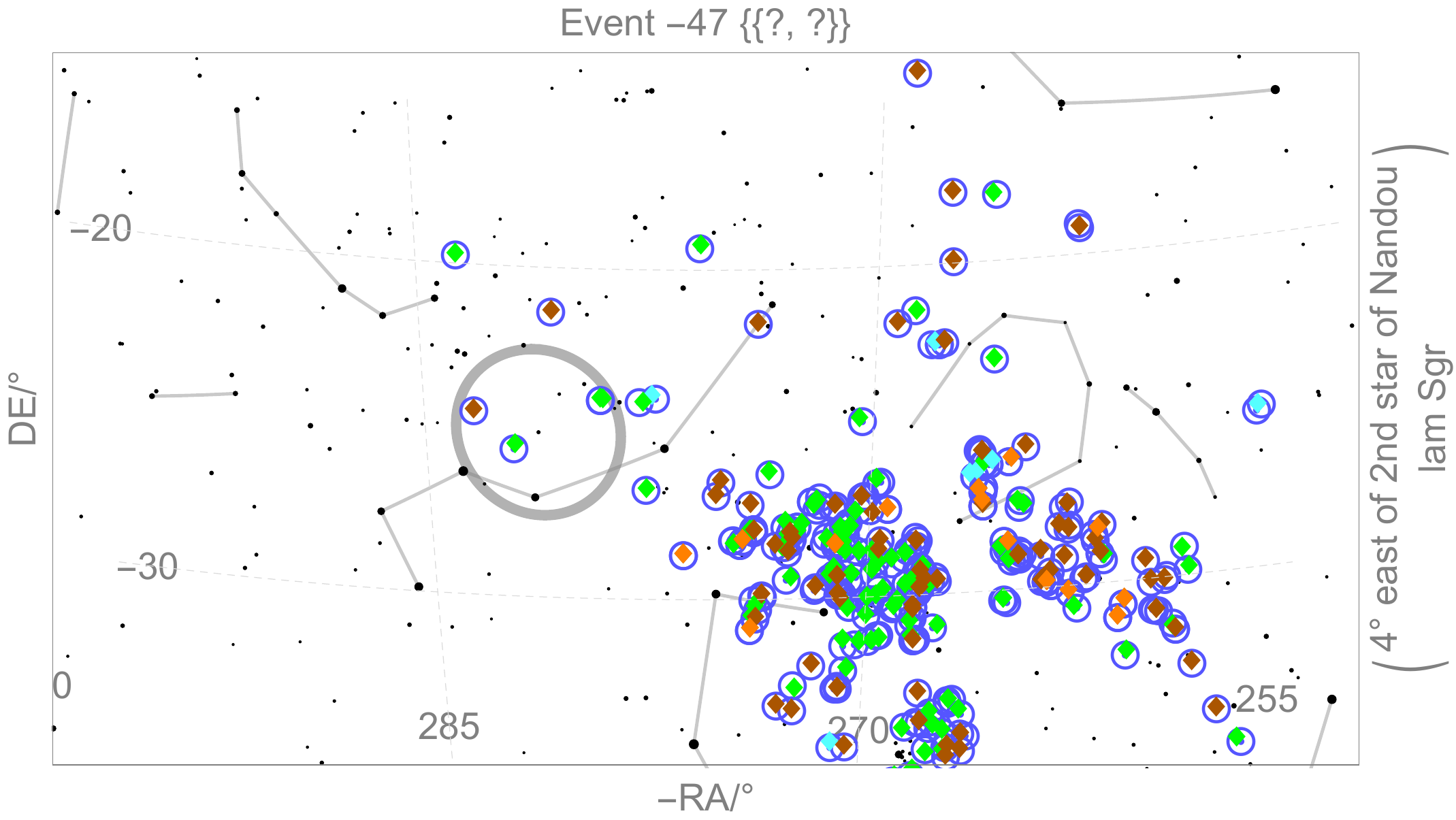} \\
	\includegraphics[width=.88\columnwidth]{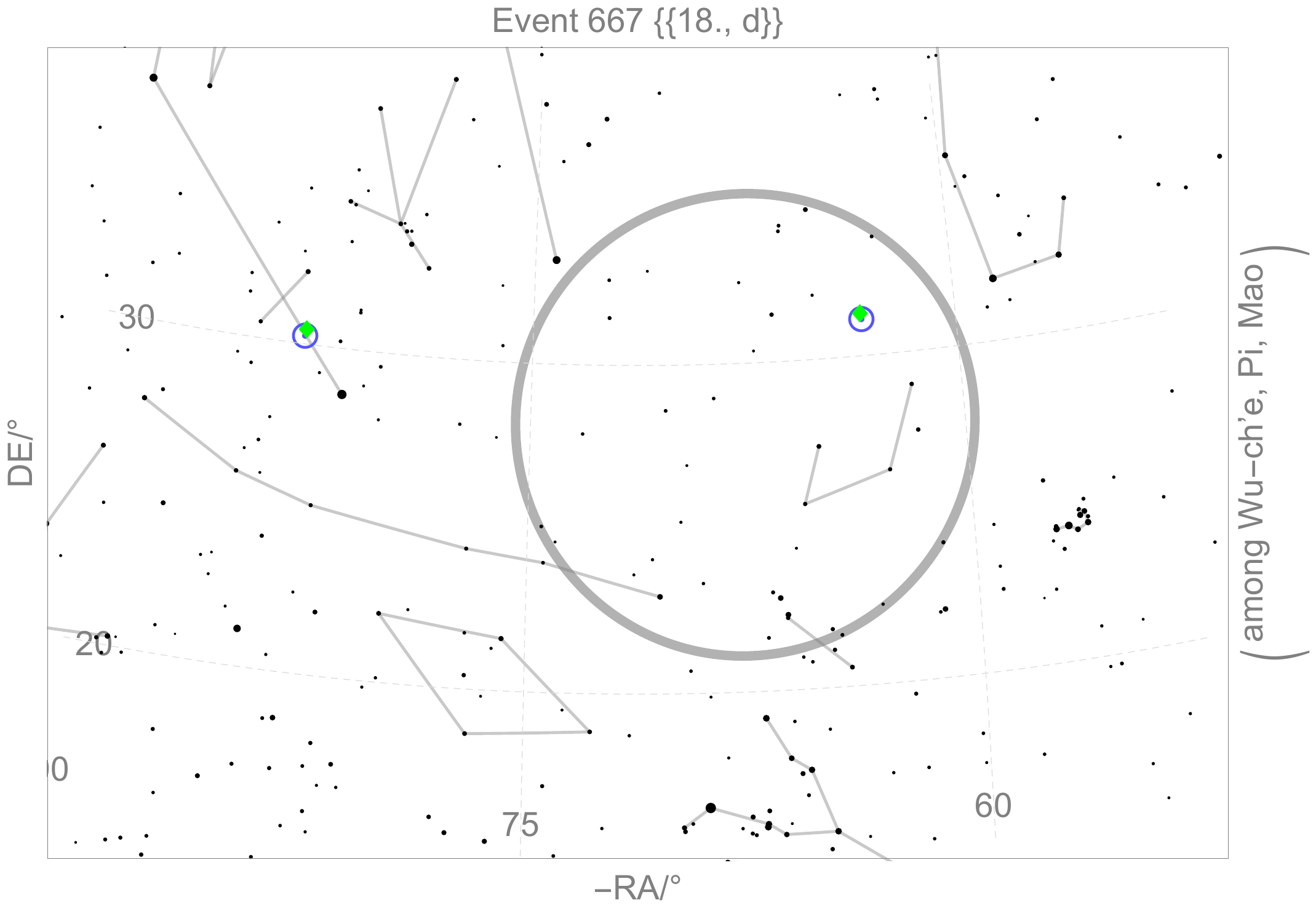} \\
	\includegraphics[width=.88\columnwidth]{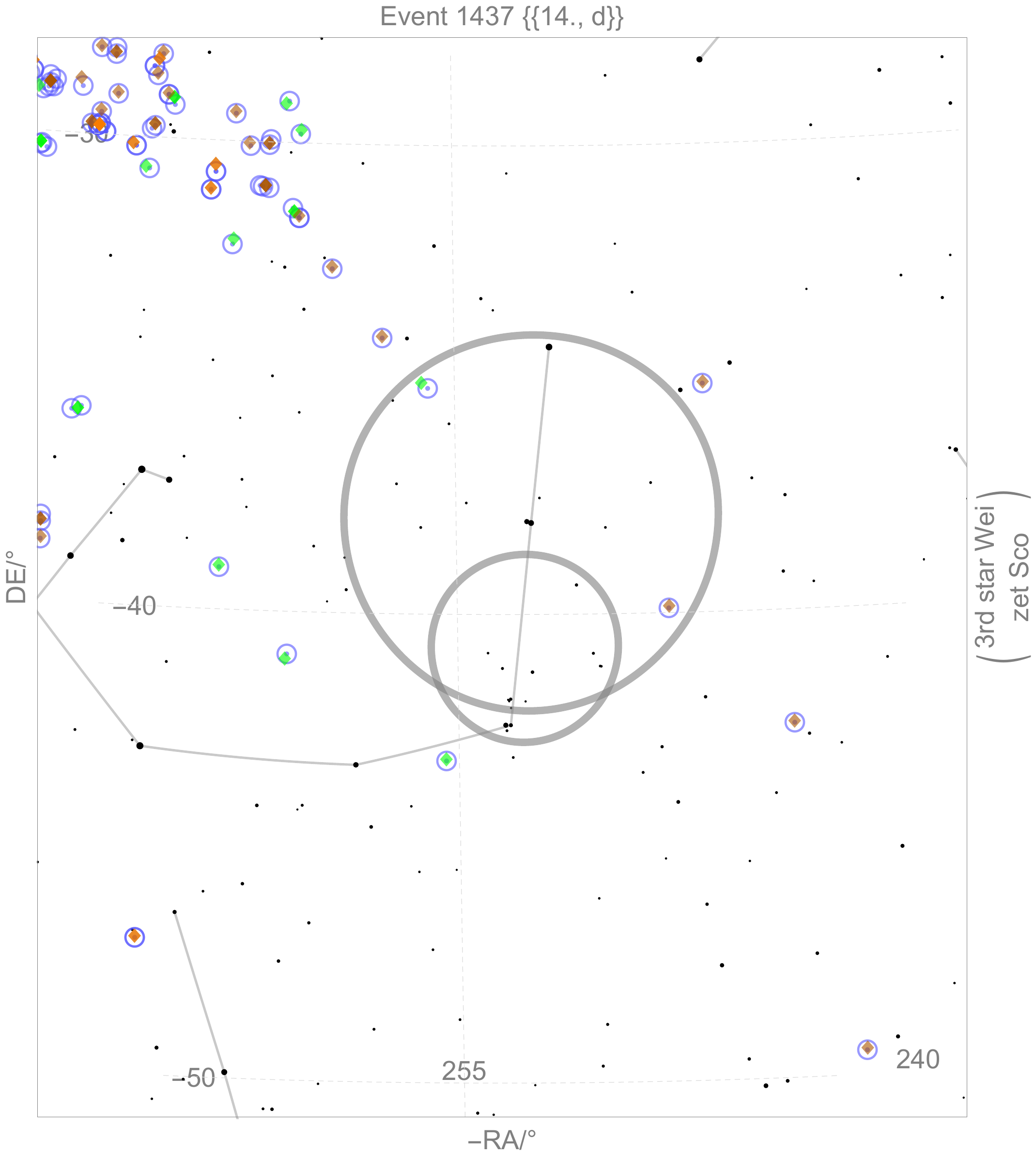}
\end{figure}

 There are only seven close pairs in our search fields identified by our routine (Tab.~\ref{tab:pairs}). The corresponding maps are displayed in Fig.~\ref{fig:pairs}. 
 
  \textbf{Close pairs in circle of event 1437:} Looking at the map (Fig.~\ref{fig:pairs}, bottom), both pairs are too far from the line between the second and the third star of Wei. 

Nevertheless, we should briefly comment on the two close pairs: Both nebulae are classified as `\textit{possible} PN'. 

G349.7+04.0 has an extension of only $3\farcs5\times4\farcs0$. As the angular separation to the CV is $\sim9\farcm5$ the star is much too far away to be in any physical relation with the nebula. In the MASH catalog \citep{mash2006} the nebula is suggested a `WR\,CSPN', a planetary nebula with central Wolf-Rayet star because the spectrum shows some WR features in blue. There are also broad emission lines but as most Wolf-Rayet central stars are of WC-type and this is certainly not a WC-type spectrum it appears strange. In principle WN stars are not excluded but in this case, no central star has yet been found and the nature of the nebula is unclear. We find this to be an extraordinarily interesting and rarely studied object but certainly not a PN or nova shell and, thus, beyond the scope of this study.  

The nearby symbiotic binary V1535 Sco with a typical G-magnitude near 15.2~mag produced a nova in 2015 brightening it to a visual magnitude of 9.2~mag (8.1~mag in $I$),\footnote{\url{https://www.aavso.org/aavso-alert-notice-508}.} below the detection limit of naked-eye observers even if Nr peaks vary by 3~magnitudes. This star is located in a region close to the Milky Way and close to the horizon for observers of the northern hemisphere, so they would not be able to realise a faint (6~mag) `new star'. 

 However, both objects of this pair are interesting cases for further studies: The nature of G349.7+04.0 is unclear and V1535 Sco is possibly a WD+K-systems in a crowded area with a known 2MASS-object only $5\arcsec$ away which needs clarification.
  
  The second pair found by our routine is based on a wrong classification in Simbad: Although WRAY~15-1518 is already clarified as `not a PN' (Simbad), it is still listed as `possible PN'. WRAY 15-1518 equals Hen 2-173, a symbiotic star in \citet{allen1984}. 
  
 \textbf{Close pairs in circle of event 667:} The only close pair in our search circle is the A1-type star HD~281789 and the cataclysmic variable 2MASS J04132921+3116279 with a distance of 6\arcmin. The star is erroneously catalogued as PN candidate; a young A1 star does not have a planetary nebula or nova shell. It has a parallax of ($2.9950\pm0.051$)~mas \citep{brown} while the apparently nearby CV 2MASS~J04132921+3116279 is at a much higher distance to Earth ($\pi=0.7331\pm0.083$~mas according to ASAS-SN \citep{shappee2014}). These two objects are certainly not related: However, 2MASS J04132921+3116279 is a hardly studied CV with Balmer emission lines and a visual magnitude of 16.3 to 18.1~mag. This makes the object a candidate to produce a naked eye classical nova.  
 
 \textbf{Close pairs in circle of event $-$47:} For this event, we find three PNe in this field which have a nearby common envelope binary; for one of these PNe (G009.8-07.5) there are even two CVs at close distance. This search field is close to the Galactic centre and provides, thus, a high density of any type of object.

 PN~G009.8-07.5 is in the globular cluster M22 (10~kly distance) and only 3~arcsec in diametre \citep{gillett1989}.\footnote{\url{http://www.messier.seds.org/more/m022_pn.html}.} This would mean the event has happened at a huge distance and \citet{goettgens2019} already suggested such a scenario. For our study it is not important how far away a CV is. The only criterion is the ability to reach naked eye visibility by a classical nova eruption with a typical amplitude of 11 to 13~mag. Both dwarf novae were discovered by the OGLE survey. They have angular separations to the PN of 0\farcm8 and 4\farcm6.  The event OGLE-BLG-DN-1057 is listed in ASAS as ASASSN-V J183624.06-235429.2 which is classified as dwarf nova (UG type) at a distance of 3065~pc (9995~ly) which also suggests a cluster membership. The other event OGLE-BLG-DN-1056 is reported in ASAS database with a distance of 2613~pc (8522~ly) which is closer but still within the globular cluster. However, both dwarf novae are farther away from the centre of the nebula than the nebula is extended (by a factor of 16 or 92, respectively) which makes both pairs unlikely related. 

 The pair PN M 3-33 and V0522 Sgr consists of a planetary nebula and a dwarf nova at an angular separation of $\sim 7\arcmin$, the distances match within the error bars (parallaxes ($0.3404\pm0.098$) and ($0.2671\pm0.0762$), resp.). The nebula only extended over $0\farcm163^2$ and was detected in the near infrared. Thus, the observed angular separation of the CV is the 42.6 fold of the diametre. The variable V0522 Sgr is identified with the 17~mag star 2MASS J18480046-2522219 which is probably wrong: V0522~Sgr refers to the transient on August 16th in 1931 which brightened up to 12.9~mag. \citet{woudt2002} already suggested a 19.7~mag object 6\arcsec\ south of the 2MASS-star as counterpart for this eruption but resumed that this still leads to an amplitude of $\sim6.8$, which is still compatible with a long-interval outbursting dwarf nova of the WZ Sge type. That means, V522 Sgr is a dwarf nova which permits occasional super outbursts. Due to models of binary evolution which suggest a long quiescence after a nova eruption, the scenario would then be: The nova eruption $\sim2000$~years ago resulted in a quiescence and a brightness compared to today's dwarf nova outburst brightness. After a while, the system started again to permit dwarf nova eruptions or even awake as classical nova as observed by \citet{mroz2016}. Although it cannot be excluded that V522~Sgr produced a classical nova $\sim2000$ years ago but even such a scenario would not take this star to the observed negative magnitudes. Thus, it is neither a candidate for the cause of this appearance nor related to the above mentioned PN. 

 The object EM$^\ast$ AS 327 had been classified as `possible planetary nebula' by \citet{perek1967} but in the meantime it turned out to be the emission nebula of the Z And-type star AS~327 with a period of 823~days. The classification as PN is obviously wrong. With a $V$ magnitude of $12.6 - 13.5$~mag and located in a area of the Milky Way this star could cause a naked eye nova of 2 or 3~mag.  
 
 In sum, none of the close nebula-variable pairs matches the conditions to be able to brighten up to visible magnitudes. The CVs without nebula presented in our earlier publication \citep{hovoMNRAS2020} remain.   

 \subsection{Discussion of the individual events}\label{chap:disc}
 Of special interest is  the \textbf{record of $-$47 (48 BCE)} which reports a really bright appearance in the constellation of Sagittarius. \citet{goettgens2019} suggested a newly discovered nebula in the nearby M22 star cluster to be the remnant of this observation but catalogued this nebula to be possibly a nova shell or a planetary nebula. However, the extinction of peak brightness within the Milky Way and the reported great brightness of the historical appearance in addition to the much too young the kinematic age of the nebula leave doubts on this interpretation \citep{hoffmann2019}. The record of this event reports a rather bright sighting: `as large as a melon' and `bluish-white', which is why we consider it of negative magnitude. If this historical event had been a supernova, we should look for a pulsar or SNR in the reported area but there is only one pulsar at the edge of the search field (PSR J1836-2354A) which is much too old ($\tau_c\sim2.3\cdot10^{10}$~years). Among the 5 planetary nebula candidates, there is the symbiotic star EM$^\ast$ AS 327 (see above). A classical nova from this system might be visible but it is uncertain if it would be described as given. The four remaining candidates are PN M~3-32, PN M~3-33, PN H~2-48, and PTB~32. None of them provides a close CV and would, thus, be a strong candidate. 
 
 More interesting is the symbiotic binary V5569~Sgr which normally varies the range of $9.8 - 12.1$~mag in V. If this brightens by $\sim10$~mag it could reach at least 0~mag. Although the position and brightness do not perfectly fit (we expect the object a bit more north and the apparent magnitude negative), within the error bars this could be a candidate. 

 In the field of \textbf{the $-$4 sighting} we did not identify any valid candidate among the CVs and X-ray binaries \citep{hovoMNRAS2020}, and there is no PN and no SNR but one pulsar: PSR~J2010-1323. According to the pulsar database \citep{atnf}\footnote{\url{https://www.atnf.csiro.au/people/pulsar/psrcat/}.} this object has a characteristic age $\tau_c= 17$~Ga and is, therefore, a MSP. The wording of the historical record is rather brief: 'a \textit{hui} appeared (\dots) for 70 days' and uses the term `hui' (broom star). Hence, we conclude that this record, suggested as possible nova or supernova candidate in the lists of \citet{steph77} (and derivatives, citations) as well as \citet{hsi} because no movement is reported, possibly still refers to a comet and not to a bright nova or supernova.  
 
 \textbf{The event in 70 CE}. For this historical event, only the constellation is given and in this case we deal with one of the large constellations (named Xuanyuan, the Yellow Emperor) extending from the feet of Ursa Major to the heart of Leo ($>512\degr^2$, cf. \citep[Tab.\,3]{hovoMNRAS2020}). In this huge array, we found 13 CVs with 5 being brighter than 16~mag and have no further criterion to give one or some of them a higher likelihood than the others \citep[Tab.\,6]{hovoMNRAS2020}. With the present study (Tab.~\ref{tab:pn} and \ref{tab:snr+psr}), we add 5 PNe but none with a close binary, 1 SNR, and 14 PSRs in the field -- and no criterion to prefer one scenario. The `guest star' which `emerged in Xuanyuan for 48 days' is not described further in the record and, therefore, we cannot derive any qualities unless the duration of 48~days suggests a classical nova in preference of a supernova. Thus, we recommend to stick to the list of CVs of \citet{hovoMNRAS2020}. 

For the observation \textbf{reported from 329}, one object precisely fits the position described in the text: The young planetary nebula M97 (the Owl Nebula). There is no known close binary in the vicinity. The nebula consists of an elongated inner (cylindric) shell and a circular outer, multi-shelled part \citep{sabbadin1985, guerrero2003} and has a diametre of 2 to 3 light years. The distribution of expansion rates shows that the inner part moves slower than the outer part. The age estimate suggests several thousand years, e.\,g. $\sim7,000$~years from typical expansion rates of $42$~km\,s$^{-1}$ and the more careful determination by \citet[6,000~years]{sabbadin1985}. Additionally, there is a halo of red giant wind around the PN which has a dynamical age of $>40,000$~years \citep{manchado1992}. In other words, the oldest layers are too old and the circular shape is from neolithic epochs, an age from which circular enclosures may suggest systematical astronomical observations by humans but without textual witnesses because script was invented only in the $-3$rd millennium. 
 
 Due to the perfect match of the positions of M97 and the guest star, we considered PN re-birth scenarios. In known cases, PN Abell~30 and PN Abell~78, the central area of the nebula is almost free from hydrogen which is not the case for M97. The only X-ray observation of the Owl by ROSAT explains a point source radiation as background \citep{chu1998} and the UV spectrum does not show any features of stellar wind. According to \citet{guerrero2003} the innermost cavity of the nebula could possibly even collapse as a consequence to this drop or lack of stellar wind. A DIN is also no option because the guest star was visible only 23~days and not many years. Apparently, this is a rather normal white dwarf CSPN \citep{mccarthy1997} in a normal PN. Consequently, we can currently not identify any known scenario of possible brightening in a PN with M97. 

 \textbf{The event in 641}. There are several planetary nebulae detected in the face-on S0 galaxy NGC~4382. Even if some of these PN are wrongly classified nova shells, they are much too far ($\sim10$~Mpc) away to have caused a naked-eye event. The only pulsar in the field is PSR J1238+21 is a rather slow ($P\sim1.11836$~s) one indicating an age of the order of $10^7$~years \citep{hobbs2004}. Hence, among these alternative objects there is no further candidate and the most likely scenario remains a nova eruption of the (now) nova-like VY Sculptoris star SDSS J122405.58+184102.7 as suggested in \citet{hovoMNRAS2020}. 
 
  \textbf{The event in 667}  There is no SNR in the field but three PSRs which are relatively old but not MSPs. The `possible PN' 2MASS J04595608+2706020 is indeed a Seyfert galaxy and, thus, accidentally in the list of PN candidates. Of the other six possible PNe in the field, two (namely V$^\ast$ V1023~Tau and CoKu Tau-Aur Star~4) are likely not PNe but T~Tauri stars (young). They could show some variability but do not brighten enough for twilight visibility by naked eye observers. The only close pair containing the A1-type star HD 281789 turned out to be a projection because the different distances of the components (see above section) and the A-type star is not a PN. PN K~3-66 is a young planetary nebula with a clearly detected dust lane\footnote{see Compilation IV: Comparison Sample: Young pPNe \url{http://faculty.washington.edu/balick/pPNe/}.} and, thus, not a mis-classified nova shell. PN K~3-29 was observed in the Infrared campaign by \citet{stanghellini2012} to analyze the dust in the nebula. The results also suggest a PN and not a nova shell and no central binary. 

 The only interesting object is NGC~1514, a PN with a size of $(1\farcm673)^2$ and binary central star HD~281679 (sdO+A0/3III C), according to \citet{feibelman1997} and citations therein. The subdwarf O-type star seems to cause a stellar wind which is visible in form of weak P-Cygni profiles of the O\,{\sc v} line at 1371~\AA\ and the C\,{\sc iv} doublet at 1549~\AA. The nebula is, thus, certainly not a nova shell but formed by the stellar wind. 

 Summarizing, none of the objects in the field of our search for possible causes of the historical sighting can be definitely identified as a counterpart. 
  
 For the \textbf{sighting in 668}, there is only one PN slightly outside our search field: NGC~2242, a very blue nebula with a diametre of $0\farcm37$ and a distance of $\sim2000$~pc. If this was a nova shell, it would be much too old (5,000~yr) to have caused a sighting only $\sim1.3$ thousand years ago; it is likely a real planetary nebula. More interesting are the supernova remnants in the field: SNR~G160.8+02.6$=$G160.9+02.6 has an estimated age of 4,000 to 7,000~years while its pulsar's $\tau_c$ is estimated to be 16,000~years (U Manitoba catalogue). The possibly related pulsar is one of the rare X-ray pulsars, the magnetar SGR 0501+4516. 

 As the phenomenon in year 668 was only visible in twilight, we already suggested that it was probably rather bright. This makes the observation of a supernova more likely than one of the six CV candidates which we found in the search field \citep[Tab.\,6]{hovoMNRAS2020}. In contrast, the object in year 668 is reported only several days (maximum three weeks) which would be atypical for a bright supernova and preferably suggests a comet or a classical nova. However, in comparison with the last paragraph on the event in $667$, something else appears really strange: In two subsequent years, there are two comets (or novae) at almost the same position in the sky (close to the Wuche-asterism). A writing error seems unlikely in this case, because not only the year is given but two different reign periods. 

 \textbf{A new hypothesis:} Interestingly, the record of $667$ only reports an appearance on 667 May 24th while the record of $668$ only reports a disappearance in the beginning of June.\footnote{\citet{hsi} understands the text to report a duration, while \citet{ho} translates the record with an addition as if it was reporting a date of disappearance; in both ways, the astronomical conclusion would be an invisibility either from 668 June 7th or latest from June 14th on.} Could this mean that the appearance of May 667 and the disappearance of June 668 refer to the same transient? 

In this case, a year of visibility would be unreported which is possible because astronomical diaries are not preserved. Additionally, we have to re-interpret the position given for the sighting in $667$: We suggest, it was originally not meant to have occurred between three asterisms (which would be, indeed, a very unusual description) but next to the Wuche asterism (as event $668$). This is (automatically) in the area of the two lunar mansions Mao and Bi which are also mentioned in the text for $667$. The different descriptions of the positions explain why up to now nobody brought these sightings together. However, we consider it more likely that the two records together report one supernova than that there were two comets at almost the same position. 

 Given a supernova in 667/ 668, we have to consider all five SNRs in our search field in more detail: $(i)$ Although age and distance of SNR G159.6+07.3 are unknown, according to \citet{fesen2010} this is not the young remnant of a historic supernova. They consider it either very old or a few thousand years old (pre-historic), if it expands in a very low density interstellar medium (ISM) resulting in a high shock velocity suppressing the formation of common postshock optical cooling filaments. $(ii)$ SNR G156.4-01.2 $=$ CTB 13 is neither in the U Manitoba nor in the Green catalogue of SNRs. Optical Observations by \citet{rosado1982} suggest a diametre of 30 to 60~pc which makes it too large for a historical event. $(iii)$ In case of the radio source G160.1-01.1 it is not even certain that this is a SNR; it is classified as `uncertain SNR' in the SNR catalogue. We are left with the two SNRs $(iv)$ SNR G160.9+02.6 and $(v)$ SNR G156.2+05.7 of which the first is the younger one. $(v)$ SNR G156.2+05.7 is estimated to be 7,000 to 26,000~years old and has no known pulsar \citep{katsuda2016} while $(iv)$ SNR G160.9+02.6 is considered to be a few thousand years younger (see above paragraph). With our new suggestion we have to study this in more detail: \citet{leahy2007} used radial velocity measurements of H\,{\sc i} filaments to infer a distance to the SNR of $0.8\pm0.4$~kpc. By using an evaporative cloud model, they obtained the above mentioned age estimate of $4000-7000$~yr for $d = 0.8$~kpc, without taking into account the distance uncertainty. At $d=0.8$~kpc and for a diameter of $130\arcmin\times120\arcmin$, the radius would be 15~pc, which is too large to be the remnant of a historical supernova. However, for smaller distances, the radius and age would shrink correspondingly.  

 Consider the two neutron stars in this SNR: ($i$) PSR J0502+4654$=$B0458+46 is located $\sim17\arcmin$ to the northeast of the geometric centre. According to the ATNF pulsar database \citep{atnf} it has a distance of 1.32~kpc, which could be consistent with the SNR distance. The characteristic age of $\tau_c=1.81\times10^6$~yr, however, seems very high. Additionally, the proper motion of PSR J0502+4654 does not point back to the geometric centre. Only if the explosion happened significantly off the centre ($\gtrsim10\arcmin$), we could still justify an association. The SNR catalogue \citep{manitoba} states this pulsar to be probably unrelated. ($ii$) In contrast, the magnetar SGR 0501+4516 is given as possibly related but with $\sim1.38\degr$ to the south it is much further away from the centre. With $\tau_c=1.6\times10^4$~yr it is relatively young but the distance of 2.2~kpc is larger and if the explosion happened close the geometric centre, it would have required an unreasonably high velocity of more than $3400~\text{km\,s}^{-1}$ to travel that far in 5,500 years (upper age limit). This seems impossible compared to the Maxwellian distribution of neutron star velocities which peaks at $265~\text{km\,s}^{-1}$ \citep{hobbs2005}. If SGR 0501+4516 was the associated neutron star, it would certainly rule out an association with the transient observed in 667/ 668. 

Both neutron star associations seem unlikely and an association of the SNR shell with the transient would challenge the evaporative cloud model used by \citet{leahy2007}, requiring a very fast expansion of the ejecta. However, using the lower distance limit of 0.4~kpc should decrease the lower age limit to about 2000~yr, i.\,e. only slightly larger than required. Summarizing, if SNR G160.9+02.6 is the remnant of the SN~667/668, definitely none of the two known PSRs are related to it and the lowest possible age limit applies. Alternatively, the remnant of this suggested new SN is not yet discovered. 

  \textbf{The event in 683.} For this event we have two objects in the field: The huge but faint supernova remnant SNR G159.6+07.3 extends $3\times4$~degree and the H$\alpha$ line emissions show velocities of $\sim170$~km\,s$^{-1}$ \citep{fesen2010} which excludes it as cause of a sighting in 683. The other object is the planetary nebula IC~2149 with an age of 350 thousand years. Would this be a nova shell it still had an age of 12 to 30 millennia and, thus, too old to be related to the event in 683. Of the four CVs we presented in \citet{hovoMNRAS2020}, three are likely too faint and the only remaining candidate is the dwarf nova ASASSN-14gy with a brightness varying between $15.74$ and $<17.4$~mag.
  
   \textbf{The event in 722} appeared in the Chinese constellation of Gedao. In this huge search field in the Cassiopeia-region of the Milky Way, we found a total of 24 planetary nebulae, 15 pulsars and even 5 supernova remnants. Most interesting to mention from the latter two object classes is the pair of the young SNR G130.7+03.1, also known as SNR 3C~58 and the young pulsar PSR~J0205+6449. Although the characteristic age of the pulsar is estimated of the order of 5,370~years, this pair has been suggested \citep{kothes2013} and is commonly treated as a candidate to have caused the guest star in 1181 which, thus, would have been a supernova. Within the margin of error of the age estimation it could also be well possible a supernova in 722 instead of 1181. However, the observation from Japan in 722 is reported only `for a total of 5~days' \citet[p.\,135]{xu2000} which makes it less likely to be a supernova. As it appeared within the Milky Way, i.\,e. was observable and recognised naked eye against a bright celestial background of many stars, we think it should have been a rather bright appearance (2.5~mag minimum, likely brighter). There is only one record from Japan, none from China or Korea, which suggests that the bright object really was invisible after a few days and the continental colleagues could have missed it for weather reasons. Yet, fading from this brightness to invisibility (4~mag) in only 5~days would also be an extremely fast nova. A supernova would probably have lasted longer and the guest star in 1181 indeed was observed from China and Japan, and it was reported for 156 or even 185 days in China. That means, the interpretation of the appearance in 1181 as supernova is still more likely than the one of $722$. Still, the event in $722$ has 9 remaining CV candidates of which one (HT Cas) had been suggested by \citet{duerbeck1993} to have caused the sighting. We add the symbiotic binary V0832 Cas with a normal range from $11.66 - 13.5$~mag in V. If the sighting was caused by a very fast nova with peak brightness of $\leq2$~mag this system is a valid candidate. 
   
In the search field of \textbf{event 840} we found two objects in the catalogue of planetary nebulae: RAFGL~3068 and PN Jn~1. The first nebula has a central carbon star (LL Peg) which shows variability of Mira type. Thus, the surrounding nebula is correctly classified as PN G093.5-40.3, implying that the object RAFGL~3068 is certainly a PN and not a nova shell. The other PN, JN~1 is discussed as `old PN' by \citet{napiwotzki1995}. It has a kinematic age of the order of 900,000~years and, thus, would be much too old for historical timescales even if we applied the higher, nova-typical expansion rates. There are no SNR in this area and among the eight PSRs three are MSPs. PSR J2355+2246, PSR J2329+16, PSR J0006+1834, PSR B2315+21, and PSR J2317+29 remain possible. 

  \textbf{The event in 891.} In this search field there is no PN but one old pulsar. In \citet{hovoMNRAS2020}, we already found the recurrent nova U~Sco in the search field. U~Sco is the most active recurrent nova known, with an eruption cycle of $10-15$~years, reaching at maximum light V$=7.5$~mag. Both objects, U~Sco and the pulsar PSR J1633$-$2010 are close to the asterism and within a circle around the mentioned star ($\psi$~Oph). However, we had to apply an error bar of $\geq3\degr$ as derived from historical supernovae \citep{hoffmannVogtProtte} to obtain candidates in our search circle. In contrast, the guest star record from Japan states an angular separation of $0\fdg1$ (one `cun') to this star. U~Sco is a little bit more than 2\degr\ away from the star. We have to postulate a corrupted text preserving 1~cun instead of 1~chin ($\sim1\degr$) because in cases of a given angular separations from a certain star (cf. SN~1054) the ancient measurements were rather precise. No candidate is found within $0\fdg1$ of $\psi$~Oph, however there are some within a 4\degr\ error circle \citep{hovoMNRAS2020}.  
  
The \textbf{event in 1175} led to no suggestion of the CVs in the field \citep{hovoMNRAS2020}. Now, we add one planetary nebula: PG 1520+525, also known as Jacoby~1. There is a bright foreground star (HD~137000) at a distance of only 3\farcm3. This spherical object is $11\times11$~arcmin$^2$ in diametre which translates to a physical extension of (3.2~pc)$^2$ \citep{tweedy1996} or a radius of $\sim1.6$~pc. With typical expansion rates of planetary nebulae (42~km\,s$^{-1}$, \citet{jacob2013}), this leads to a kinematic age of the order of $4\cdot10^4$~years for the PN. With the typical expansion rate of nova shells, the nebula is only 1,300 to 3,200 years old and, thus, the kinematic age fitted the event in 1175. However, no CV is yet known in the vicinity (no suggestion in \citet{hovoMNRAS2020}) and there is no known common envelope-binary inside the nebula. 

For the \textbf{event 1430}, the only object in the field is a planetary nebula from the Hartl-Dengel-Weinberg (HDW) catalogue: PN HDW~7. The object has an extremely blue central star which already had been suggested as CV \citep{gentile2015}. However, its radius of $0\farcm8$ translates to $~57$~pc at a distance of 4.3~kpc. With a bulk velocity determined from H$\alpha$ measurements \citep{pereyra2013} of $21$~km/s it has an age of the order of $2.5\cdot10^{6}$~years and is, therefore, too old to have been caused in 1430. Still, there could be a CV within the observable (faint) nebula which did not cause the whole nebula but a fainter one within the visible nebula. The possibility that the observed object is the remnant of a sun-like star which resulted in a white dwarf a million years ago while the resulting white dwarf produced a nova in 1430 is not yet excluded. However, our search for close binaries within this nebula (or even the search field) did not result in a match for this event: There is no currently known CV or symbiotic system in the vicinity. 

  \textbf{The event in 1431.} Of the three objects listed as `possible PN' in this field, the Witch Head Nebula (NGC~1909) is a reflection nebula. Because of its size $3\degr\times1\degr$ and distance (1,000~ly) it cannot be the remnant of an event only $\sim600$~years ago. In the planetary nebula PN G095.2+00.7 (PN A66 73), an unusual carbon distribution was found \citep{ohsawa2012} and PN MaC~2-1 (PN G205.8-26.7) is a PN at a distance of more than 9~kpc \citep{stanghellini2016}. The five PSRs in the field have characteristic ages of the order of $10^6$ to $10^7$~years and are, thus, rather old pulsars.

 On the other hand, the Z~And-type star KT~Eri has erupted as a classical nova in 2009 reaching a visual maximum brightness of at least 5.4~mag in the regime of naked-eye visibility. KT~Eri is an excellent candidate for a recurrent nova. As already mentioned in Section~\ref{chap:methZAND}, the RS~Oph group among recurrent novae contain red giant secondaries, like Z~And stars. The rather faint quiescent magnitude V$ = 16.6$ corresponds to the eclipse minima of KT~Eri, outside eclipses the star is around 15~mag, according to the light curve in the AAVSO Light Curve Generator. Therefore, its nova eruption amplitude of $\sim10$~mag is well comparable to that of modern recurrent novae. The historical records mentions `shiny bright' for the event of 1431, implying a rather bright guest star but they also list a visibility duration of only 15 days \citep[Tab.\,3]{hoffmannVogtProtte} compatible with the modern classification of KT~Eri as a rapid Na-type nova. All this opens the fascinating perspective of having perhaps identified a recurrent nova with a cycle of $\sim600$~years, while all cycles of modern recurrent novae are shorter than $100$~years. Additionally, as the peak magnitude of the nova eruption in 2009 was 5.4~mag, the hypothesis of an outburst of this system $\sim600$~years ago as `shiny bright' would imply a much bigger peak-to-peak variability of nova outbursts in a binary system as observed until now: e.\,g. in case of T~Pyx the peak brightness varies by 1~mag \citep{mayall1967,schaefer2010} and for U~Sco the peak brightness varies by 2 to 3 mag (see AAVSO light curve archive). If an outburst $\sim600$~years ago would have reached 3~mag or even 2~mag it would be easily recognizable for naked eye observers and they would call it `bright' in regions like constellations Psc or Her -- but in a region close to the local horizon, next to the Milky Way and next to some of the brightest stars in the sky (CMa-Ori-region) they would call something `shiny bright' only if it was about 0~mag or brighter, say comparable with Sirius ($-1.46$~mag). This makes the suggestion of KT~Eri especially controversial. It would be an interesting case but there are 8 further CVs in the field of this event which could be alternative suggestions \citep[dwarf nova BF Eri and the dwarf nova of SU UMa-type AQ Eri as brightest ones]{hovoMNRAS2020}. None of them offers such unique and seldom properties as KT~Eri. 
 
 Especially interesting is, of course, the \textbf{event in 1437} because of the famous suggestion of \citet{shara2017_nov1437}. As it is located close to the centre of the Milky Way, the density of any types of objects in this field is rather high. There is no SNR but there are 27 PSRs and 51 planetary nebulae in the search circle (defined for the query in the VSX \citep{watson}). However, in this case, we know that the position should be (more or less) at a line connecting two stars from $\zeta$~Sco northwards. Thus, we can neglect the objects at the eastern and western edge of the search circle. Still, there are 19 PNe and 8 PSRs to consider: The best positioned PSRs are PSR J1654-4140, PSR J1650-4126, and PSR J1653-4030 but two of them are MSPs. The remaining PSR with $\tau_c\sim10^7$~years appears also too old although the true age of a pulsar is younger than the characteristic age. 

 Among the many planetary nebulae in the field, only four objects have the position matching the description: PN PM~1-114, PN G344.4+01.8, PN G344.0+02.5, and PN G344.2+01.6. None of these objects contains a known common envelope binary found by routine. The two close pairs within the search circle (Tab.~\ref{tab:pairs}) are too far from the line from $\zeta$~Sco northwards and, thus, are no candidates. PM~1-114 has a huge distance (plx$\sim0.0074$~mas, \citet{brown}) and an uncertain PN. G344.4+01.8 and G344.0+02.5 are enlisted in the MASH\,II planetary nebula catalogue but only their size $9\arcsec$ is known and if they are PNe is only `likely'. The G344.2+01.6 was only found by the AKARI mid-infrared survey as pale red dot without measured size or distance. These objects need further investigation to determine their real nature, not to mention any sort of rebirth scenario.  

   \textbf{The event in 1461.} There is one planetary nebula in the field: PN G048.7-01.5 (DeHt~4), no SNR, and 9 PSR. The PN has a size of roughly $44\times34\arcsec$ and a distance of $\sim5000$~pc \citep[object No.\,4 in Tab.~1]{dengel1980}. Thus, the physical radius should be roughly $\sim30$~pc and, thus, a kinematic age of the order of 650 millennia in case of a PN and 23 to 55 in case of a nova shell. Anyway, this object is much too old to be related to the event in 1461. 
 
  PSR J1738+0333 is a millisecond pulsar at a distance of 1.47~kpc, PSR J1739+0612 is listed with a rotational period of 0.23~s and a characteristic age of $2.37\cdot10^7$~years. The other PSRs are rather old pulsars with rotational periods of 0.999 to 1.9~seconds. With a visibility duration of only three or four days, this sighting is highly unlikely a supernova report. Thus, the three CVs from \citep{hovoMNRAS2020} still have a higher likelihood to have caused this event if it was a transient stellar object. 
   
  \textbf{The event in 1497.} Within the search circle of 1497 there is no PN and no SNR. There are two PSRs, PSR J1434+7257 and [KKL2015] J1439+76. The first one is a millisecond pulsar and, thus, with a characteristic age of $1.2\cdot10^{9}$~years too old to have caused a brightening in 1497. The other one also seems to be rather old with a period of almost a second \citep{atnf}.   

  \textbf{The event in 1661.} There is no PN or SNR in the field neither a CV suggested to be bright enough. The rotating variable V$^\ast$ LY~Aqr contains a MSP and is, thus, much too old for recent supernovae. Concerning the companion star, which has an orbital period of 2.38 h, \citet{lazaridis2011} argue that it is more likely to be a semidegenerate helium star than a white dwarf which makes a classical nova as unlikely as a supernova. As the sighting in 1661 is reported `as bright as Saturn' ($\sim0.5$~mag), our study does not return a suggestion. 
  
  \textbf{The event in 1690.} In the search circle of the event in 1690, there is no PSR and no SNR but 9 planetary nebulae. None of them is in the vicinity of a common envelope binary. LSE~63 is a post-AGB star and, thus, a proto-PN. IC~4776 is a nebula with a post-common envelope central binary. The components are likely a peculiar WC star \citep[peculiar because no $\lambda5806$]{allen1985} and/ or a low-mass main sequence star \citep{sowicka2017}. According to the MASH catalogue, G002.1-08.3 is a likely PN of $9\times7\arcsec$, G001.1-11.5 a possible PN of $5\times4\arcsec$, while PN G002.8-10.7 ($15\times13\arcsec$) and PHR J1833-3115 $=$ G002.9-10.2 ($25\times25\arcsec$) are a true PNe. PN SB~55 and M~3-29 are only enrolled in \citet{frew2013}'s catalogue of true PNe. \cite{pena2017} found an expansion velocity of 24~km/s for PN Cn~1-5 which confirms it as true PN ($0.213\times0.196\arcmin$) with central Wolf-Rayet star (WC). In sum, IC~4776 as ejecta of a post-common envelope binary is the most interesting case in this subset but they derive an age of the nebula of $\sim1500$~years. As producing such a nebula is a long process and since starting this process the system evolves to less activity, we do not see any process which could have caused a brightening of this system 330~years ago. From \citet{hovoMNRAS2020}, the dwarf nova V1595 Sgr remains as the brightest of three dwarf nova candidates.  

\section{Conclusions}
 In the present work we checked all currently known objects in our search fields for counterparts of selected historical transients. If none of the above mentioned object types yields a positive result in the form of a proper candidate which could have caused the sighting, it could be a comet, a yet undiscovered writing error or misunderstanding in the historical process of copying and transforming astronomical reports to chronicles, or a star in the field which is not yet discovered or which has (or had) outburst qualities which are yet unknown. However, nowadays, in the age of big data science, this is less likely than in the 19th or 20th century. Remember, the star BK~Lyn was just discovered in 1986 when \citet{hertzog1986} suggested it as candidate to explain the Chinese guest star observation in 101. Nevertheless, our method led to some results which are summarised in Table~\ref{tab:results}; this table displays the summarised candidates from all five papers. 

 \begin{table}
	\centering
	\caption{Summarizing the results from \citet{hoffmannVogtProtte,hovoMNRAS2020} and this work.}
	\label{tab:results}
	\begin{tabular}{rp{.88\columnwidth}} 
		\hline
		year	&result so far\\
		\hline
$-203$	&AB Boo as brightest of three CVs in the field, no further suggestion\\
$-103$	&no suggestion (no objects, possibly text corrupted)\\
$-47$	&possible are novae from symbiotic binaries AS\,327 and V5569~Sgr\\
$-4$	&no suggestion: comet?\\
64	& no suggestion (1 CV, V0379 Vir, too faint \citep{hovoMNRAS2020}).\\
70	&13 CVs of same likelihood, 14 old PSR, 1 SNR\\
101	&BK Lyn still most likely (of 2 CV suggestions).\\
329	&PN M97 fits the description perfectly but is too old. Remaining are 5 CVs suggested in \citet{hovoMNRAS2020}.\\
641		& most likely candidate: the NL/VY star SDSS J122405.58+184102.7 \citet{hovoMNRAS2020}.\\
667 	&appearance of a SN  \\
668 	&disappearance of SN~667 (consider SNR G160.9+02.6 (?)) \\
683		& 1 CV: ASASSN-14gy \citep{hovoMNRAS2020}, no close pairs, 1 PN, 1 PSR both too old.\\
722		& very fast nova of HT~Cas \citep{duerbeck1993} or symbiotic V0832 Cas or one of the other 8 CVs \citep{hovoMNRAS2020} \\
840		& 5 CVs \citep{hovoMNRAS2020} and 3 old PSRs \\
891		& recurrent nova candidate: U Sco $\sim2\degr$, but it is more than $0\fdg1$ away and its peak brightness should vary more than currently observed\\
1175	& extremely fast nova? no suggestion (CVs too faint) but please study if PN PG~1520+525 could be a nova shell\\
1430	& HDW\,7 the most interesting object but BG CMi also possible (very fast nova?)\\
1431	& KT Eri as classical (possibly recurrent) nova with a $\sim600$ years eruption cycle and $\sim6$~mag peak-to-peak variability; 8 other CVs (brightest:  AQ~Eri or BF~Eri); 3 PNe and 5 old PSRs; no SNR\\
1437	& no suggestion (no CV fits the position perfectly, 4 PNe fit, 8 PSRs likely too old, 1 apparent pair is symbiotic binary)\\
1461	&3 CVs of similar likelihood from \citet{hovoMNRAS2020} (6 old PSRs, 1 PN). \\
1497	& no suggestion (no CVs, no PN, no SNR, 1 MSP, 1 old PSR)\\
1661	& no suggestion (CVs too faint, no PN, no SNR, 1 MSP, 1 old PSR) \\
1690	&3 CV candidates, 9 PNe, no close pairs\\
		\hline
	\end{tabular}
\end{table}

 In some cases, we improved the likelihood that the event was either a nova or a supernova, while in some other cases we still cannot give any proper suggestion. These events should be postponed and maybe revised in some decades with new findings although there is, of course, always the possibility that we deal with comets or that the description is corrupt. The most interesting new findings in our study are: 
  \begin{enumerate} 
  \item The two Korean records of the year 1592 which had been proposed earlier as classical novae or even the supernova which led to Cas~A, are possibly corrupt descriptions of the SNe in 1572 and 1604 \citep{hovoMNRAS2020}. 
  \item The nebula found by \citet{goettgens2019} in M22 is too young to have any correlation with the guest star in $-47$ \citep{hoffmann2019}.
  \item The eruption which lead to the nova shell discovered and properly age-dated to the 15th century by \citet{shara2017_nov1437} is possibly not preserved in Far Eastern observations because the position of record 1437 is 3 or 4\degr\ away \citep{hoffmann2019}. This is likely because chroniclers needed political reasons to include an observation in their writings. However, we did not find any alternative suggestion to explain the records from 1437 (this paper). 
  \item The event in 891 could be caused by a known recurrent nova (U~Sco).
  \item The event in 1431 could be caused by KT~Eri as classical nova eruption in a symbiotic binary. This would imply a timescale of $\sim600$~years for the recurrence and a variation of amplitudes from one peak to the next by $\geq6$~mag. 
  \item The brief records of 667 and 668 likely describe the same transient which was, thus, most likely a supernova. We cannot identify a certain SNR but G160.9+02.6 is the youngest one in the search field. 
  \end{enumerate} 
  
 \textbf{By-products:} Additionally, our search led to some by-products which do not serve to explain the historical sighting in the very case (for reasons of brightness, duration or position) but are strongly recommended to observe: 
\begin{itemize} 
 \item PN HDW~7 has an extremely blue central star which already had been suggested as CV \citep{gentile2015} but the CV is uncertain. 
 \item IC~4776 is a nebula with a post-common envelope central binary, aged $\sim1500$~years.
 \item WRAY 15-1518 equaling Hen 2-173 is a bright symbiotic star off our position line but interesting to study.
 \item Is the symbiotic binary V1535 Sco, Nova 2015~Sco, a recurrent nova and could it possibly peak to naked-eye brightness? 
 \item The spectrum of G349.7+04.0 (MASH, VizieR) does not show the PN-characteristic emission lines from forbidden transitions, except of a weak low-excitation [O\,{\sc i}] $\lambda$5577. Instead, the spectrum reveals a stellar continuum with a few broad emission lines, most prominent at H$\alpha$/He\,{\sc ii}, H$\beta$/He\,{\sc ii}, He\,{\sc ii}\,4686, and N\,{\sc iii}\,4640. Hence, it resembles a Wolf-Rayet type of the nitrogen sequence (WN). Though, with a Gaia distance of about 1.6\,kpc and a Gaia RP magnitude of 16.4\,mag \citep{brown}, the object is too faint for a massive WN star. The low-excitation nebular emission mentioned above does not match to such a hot central star. Thus, the object could be one of the postulated underluminous `stripped' stars. 
\end{itemize}  

To obtain this result, we processed the current catalogues of cataclysmic variables, planetary nebulae, pulsars, and supernova remnants, e.\,g. from CDS Simbad, AAVSO VSX, U Manitoba, ATNF, and others. This is not yet the end: In order to explain the historical sightings, we will have to process all data of variable stars and galaxies in our search fields, study historical maps and star catalogues in more detail in order to increase the geographical and temporal resolution of the usage of certain standards in various science cultures (e.\,g. how stars are named, which variant of drawing a constellation was preferred in which place and at what epoch), and consider the historical astrological importance of the mentioned constellations in order to estimate whether or not and in which way a chronicler could have modified the original (and not preserved) astronomical record. As constellations in China have been much more stable than in Europe during the last $\sim2$ millennia, this will not completely falsify our present results -- but in some cases it could improve them (in some cases maybe increase the size of the search circle, in some cases vary the positions by opening further possibilities for the interpretations). We hope to be able to continue this threat in the future.

\section*{Acknowledgements}
 We thank Helge Todt and Wolf-Rainer Hamann (both U Potsdam, Germany), Oliver Lux (AIU, Friedrich-Schiller-Universität Jena), Jesse Chapman (UC Berkeley, CA, USA), David Pankenier (Lehigh University, Bethlehem, PA, USA), and Nalini Kirk (HU Berlin, Germany) for fruitful discussions and interesting questions on some details. We thank our benevolent referees, Mike Shara (American Museum of Natural History) and Virginia Trimble (UC Irvine) for their very useful comments to improve the readability of this paper and their very interesting recommendations of additional recent literature. Mark Booth (AIU, Friedrich-Schiller-Universität Jena) gave some advise on text structure and English language style. S.H. thanks Artie Hatzes (Thuringian State Observatory, Germany) for his support as mentor and in terms of English language style.\\
 S.H. thanks the Free State of Thuringia for financing the project at the Friedrich Schiller University of Jena, Germany. N.V. acknowledges financial support from FONDECYT regularNo. 1170566 and from Centro de Astrofísica, Universidad de Valparaíso, Chile. \\
Thankfully we made use of the VSX variable star catalogue of the American Association of Variable star Observers (AAVSO) \citep{watson}, of the SIMBAD data base (CDS Strassbourg) \citep{wenger2000}, the MASH and MASH\,II catalog of planetary nebulae \citep{mash2006,mashII2008}, the High energy SNRcat of the U Manitoba \citep{manitoba}, the ATNF Pulsar Catalogue \citep{atnf}, and of Stellarium \url{http://stellarium.org} \citep{stellarium}. Ralph Neuhäuser (AIU, Friedrich-Schiller-Universität Jena) had the initiative and idea to reconsider historical nova identifications in a transdisciplinary project. We thank him to have brought us together.
 



\bibliographystyle{mnras}
\renewcommand{\refname}{Papers of our Series}

\renewcommand{\refname}{References}
\bibliography{altNovaeA} 

\begin{thebibliography}{}
\makeatletter
\relax
\def\mn@urlcharsother{\let\do\@makeother \do\$\do\&\do\#\do\^\do\_\do\%\do\~}
\def\mn@doi{\begingroup\mn@urlcharsother \@ifnextchar [ {\mn@doi@}
  {\mn@doi@[]}}
\def\mn@doi@[#1]#2{\def\@tempa{#1}\ifx\@tempa\@empty \href
  {http://dx.doi.org/#2} {doi:#2}\else \href {http://dx.doi.org/#2} {#1}\fi
  \endgroup}
\def\mn@eprint#1#2{\mn@eprint@#1:#2::\@nil}
\def\mn@eprint@arXiv#1{\href {http://arxiv.org/abs/#1} {{\tt arXiv:#1}}}
\def\mn@eprint@dblp#1{\href {http://dblp.uni-trier.de/rec/bibtex/#1.xml}
  {dblp:#1}}
\def\mn@eprint@#1:#2:#3:#4\@nil{\def\@tempa {#1}\def\@tempb {#2}\def\@tempc
  {#3}\ifx \@tempc \@empty \let \@tempc \@tempb \let \@tempb \@tempa \fi \ifx
  \@tempb \@empty \def\@tempb {arXiv}\fi \@ifundefined
  {mn@eprint@\@tempb}{\@tempb:\@tempc}{\expandafter \expandafter \csname
  mn@eprint@\@tempb\endcsname \expandafter{\@tempc}}}

\bibitem[\protect\citeauthoryear{{Acero} et~al.,}{{Acero}
  et~al.}{2016}]{acero2016}
{Acero} F.,  et~al., 2016, \mn@doi [\apjs] {10.3847/0067-0049/224/1/8}, \href
  {https://ui.adsabs.harvard.edu/abs/2016ApJS..224....8A} {224, 8}

\bibitem[\protect\citeauthoryear{{Allen}}{{Allen}}{1984}]{allen1984}
{Allen} D.~A.,  1984, Proceedings of the Astronomical Society of Australia,
  \href {https://ui.adsabs.harvard.edu/abs/1984PASAu...5..369A} {5, 369}

\bibitem[\protect\citeauthoryear{{Aller} \& {Keyes}}{{Aller} \&
  {Keyes}}{1985}]{allen1985}
{Aller} L.~H.,  {Keyes} C.~D.,  1985, \mn@doi [\pasp] {10.1086/131677}, \href
  {https://ui.adsabs.harvard.edu/abs/1985PASP...97.1142A} {97, 1142}

\bibitem[\protect\citeauthoryear{{Anderson} et~al.,}{{Anderson}
  et~al.}{2017}]{anderson2017}
{Anderson} L.~D.,  et~al., 2017, \mn@doi [\aap] {10.1051/0004-6361/201731019},
  \href {https://ui.adsabs.harvard.edu/abs/2017A&A...605A..58A} {605, A58}

\bibitem[\protect\citeauthoryear{Baade}{Baade}{1943}]{baade}
Baade W.,  1943, ApJ, 97, 119

\bibitem[\protect\citeauthoryear{{Bl{\"o}cker}}{{Bl{\"o}cker}}{2003}]{bloecker2003}
{Bl{\"o}cker} T.,  2003, in {Kwok} S.,  {Dopita} M.,   {Sutherland} R.,  eds,
  IAU Symposium Vol. 209, Planetary Nebulae: Their Evolution and Role in the
  Universe. p.~101 (\mn@eprint {arXiv} {astro-ph/0207161})

\bibitem[\protect\citeauthoryear{Bode \& Evans}{Bode \& Evans}{2008}]{bode}
Bode M.~F.,  Evans A.~e.,  1989, 2008, Classical Novae.
Cambridge University Press

\bibitem[\protect\citeauthoryear{{Boffin} \& {Jones}}{{Boffin} \&
  {Jones}}{2019}]{Boffin2019}
{Boffin} H. M.~J.,  {Jones} D.,  2019, {The Importance of Binaries in the
  Formation and Evolution of Planetary Nebulae},
  \mn@doi{10.1007/978-3-030-25059-1.
}

\bibitem[\protect\citeauthoryear{{Burgay} et~al.,}{{Burgay}
  et~al.}{2006}]{burgay2006}
{Burgay} M.,  et~al., 2006, \mn@doi [\mnras]
  {10.1111/j.1365-2966.2006.10100.x}, \href
  {https://ui.adsabs.harvard.edu/abs/2006MNRAS.368..283B} {368, 283}

\bibitem[\protect\citeauthoryear{{Chu}, {Gruendl}  \& {Conway}}{{Chu}
  et~al.}{1998}]{chu1998}
{Chu} Y.-H.,  {Gruendl} R.~A.,   {Conway} G.~M.,  1998, \mn@doi [\aj]
  {10.1086/300571}, \href
  {https://ui.adsabs.harvard.edu/abs/1998AJ....116.1882C} {116, 1882}

\bibitem[\protect\citeauthoryear{Clark \& Stephenson}{Clark \&
  Stephenson}{1977}]{steph77}
Clark D.,  Stephenson F.,  1977, The Historical Supernovae.
Pergamon, Oxford

\bibitem[\protect\citeauthoryear{{Darnley}}{{Darnley}}{2019}]{darnley2019}
{Darnley} M.~J.,  2019, arXiv e-prints, \href
  {https://ui.adsabs.harvard.edu/abs/2019arXiv191213209D} {p. arXiv:1912.13209}

\bibitem[\protect\citeauthoryear{{Dengel}, {Hartl}  \& {Weinberger}}{{Dengel}
  et~al.}{1980}]{dengel1980}
{Dengel} J.,  {Hartl} H.,   {Weinberger} R.,  1980, \aap, \href
  {https://ui.adsabs.harvard.edu/abs/1980A&A....85..356D} {85, 356}

\bibitem[\protect\citeauthoryear{{Douchin}}{{Douchin}}{2015}]{douchin2015PhD}
{Douchin} D.,  2015, PhD thesis, Department of Physics and Astronomy, Macquarie
  University; Universite de Montpellier 2

\bibitem[\protect\citeauthoryear{{Douchin} et~al.,}{{Douchin}
  et~al.}{2015}]{douchin2015}
{Douchin} D.,  et~al., 2015, \mn@doi [\mnras] {10.1093/mnras/stu2700}, \href
  {https://ui.adsabs.harvard.edu/abs/2015MNRAS.448.3132D} {448, 3132}

\bibitem[\protect\citeauthoryear{{Duerbeck}}{{Duerbeck}}{1993}]{duerbeck1993}
{Duerbeck} H.~W.,  1993, in {Regev} O.,  {Shaviv} G.,  eds,  Vol. 10,
  Cataclysmic Variables and Related Physics. p.~77

\bibitem[\protect\citeauthoryear{Duerbeck}{Duerbeck}{2008}]{duerbeck}
Duerbeck H.,  2008, Novae: a historical perspective.
Cambridge University Press

\bibitem[\protect\citeauthoryear{{Feibelman}}{{Feibelman}}{1997}]{feibelman1997}
{Feibelman} W.~A.,  1997, \mn@doi [\pasp] {10.1086/133928}, \href
  {https://ui.adsabs.harvard.edu/abs/1997PASP..109..659F} {109, 659}

\bibitem[\protect\citeauthoryear{{Ferrand} \& {Safi-Harb}}{{Ferrand} \&
  {Safi-Harb}}{2012}]{manitoba}
{Ferrand} G.,  {Safi-Harb} S.,  2012, \mn@doi [Advances in Space Research]
  {10.1016/j.asr.2012.02.004}, \href
  {https://ui.adsabs.harvard.edu/abs/2012AdSpR..49.1313F} {49, 1313}

\bibitem[\protect\citeauthoryear{{Fesen} \& {Milisavljevic}}{{Fesen} \&
  {Milisavljevic}}{2010}]{fesen2010}
{Fesen} R.~A.,  {Milisavljevic} D.,  2010, \mn@doi [\aj]
  {10.1088/0004-6256/140/5/1163}, \href
  {https://ui.adsabs.harvard.edu/abs/2010AJ....140.1163F} {140, 1163}

\bibitem[\protect\citeauthoryear{{Frew}, {Boji{\v{c}}i{\'c}}  \&
  {Parker}}{{Frew} et~al.}{2013}]{frew2013}
{Frew} D.~J.,  {Boji{\v{c}}i{\'c}} I.~S.,   {Parker} Q.~A.,  2013, \mn@doi
  [\mnras] {10.1093/mnras/sts393}, \href
  {https://ui.adsabs.harvard.edu/abs/2013MNRAS.431....2F} {431, 2}

\bibitem[\protect\citeauthoryear{{Fujiwara}}{{Fujiwara}}{2003}]{fujiwara2003}
{Fujiwara} T.,  2003, in IAU General Assembly. p.~2

\bibitem[\protect\citeauthoryear{{Fujiwara} \& {Yamaoka}}{{Fujiwara} \&
  {Yamaoka}}{2005}]{fujiwara2005}
{Fujiwara} T.,  {Yamaoka} H.,  2005, Journal of Astronomical History and
  Heritage, \href {https://ui.adsabs.harvard.edu/abs/2005JAHH....8...39F} {8,
  39}

\bibitem[\protect\citeauthoryear{{Gaia Collaboration}, Brown, Vallenari  \& et
  al.}{{Gaia Collaboration} et~al.}{2018}]{brown}
{Gaia Collaboration} Brown A.~G.~A.,  Vallenari A.,   et al. 2018, A+A, 616, A1

\bibitem[\protect\citeauthoryear{{Gentile Fusillo}, {G{\"a}nsicke}  \&
  {Greiss}}{{Gentile Fusillo} et~al.}{2015}]{gentile2015}
{Gentile Fusillo} N.~P.,  {G{\"a}nsicke} B.~T.,   {Greiss} S.,  2015, \mn@doi
  [\mnras] {10.1093/mnras/stv120}, \href
  {https://ui.adsabs.harvard.edu/abs/2015MNRAS.448.2260G} {448, 2260}

\bibitem[\protect\citeauthoryear{{Gillett}, {Jacoby}, {Joyce}, {Cohen},
  {Neugebauer}, {Soifer}, {Nakajima}  \& {Matthews}}{{Gillett}
  et~al.}{1989}]{gillett1989}
{Gillett} F.~C.,  {Jacoby} G.~H.,  {Joyce} R.~R.,  {Cohen} J.~G.,  {Neugebauer}
  G.,  {Soifer} B.~T.,  {Nakajima} T.,   {Matthews} K.,  1989, \mn@doi [\apj]
  {10.1086/167241}, \href
  {https://ui.adsabs.harvard.edu/abs/1989ApJ...338..862G} {338, 862}

\bibitem[\protect\citeauthoryear{{G{\"o}ttgens} et~al.,}{{G{\"o}ttgens}
  et~al.}{2019}]{goettgens2019}
{G{\"o}ttgens} F.,  et~al., 2019, \aap, 626, A69

\bibitem[\protect\citeauthoryear{{Green}}{{Green}}{1988}]{green1988}
{Green} D.~A.,  1988, \mn@doi [\apss] {10.1007/BF00646462}, \href
  {https://ui.adsabs.harvard.edu/abs/1988Ap&SS.148....3G} {148, 3}

\bibitem[\protect\citeauthoryear{{Green}}{{Green}}{2019}]{green2019}
{Green} D.~A.,  2019, \mn@doi [Journal of Astrophysics and Astronomy]
  {10.1007/s12036-019-9601-6}, \href
  {https://ui.adsabs.harvard.edu/abs/2019JApA...40...36G} {40, 36}

\bibitem[\protect\citeauthoryear{{Green} \& {Stephenson}}{{Green} \&
  {Stephenson}}{2003}]{greenSteph2003}
{Green} D.~A.,  {Stephenson} F.~R.,  2003, {Historical Supernovae}.
pp 7--19, \mn@doi{10.1007/3-540-45863-8_2}

\bibitem[\protect\citeauthoryear{{Green} \& {Stephenson}}{{Green} \&
  {Stephenson}}{2017}]{greenSteph2017}
{Green} D.~A.,  {Stephenson} F.~R.,  2017, {Possible and Suggested Historical
  Supernovae in the Galaxy}.
p.~179, \mn@doi{10.1007/978-3-319-21846-5_51}

\bibitem[\protect\citeauthoryear{{Guerrero}, {Chu}, {Manchado}  \&
  {Kwitter}}{{Guerrero} et~al.}{2003}]{guerrero2003}
{Guerrero} M.~A.,  {Chu} Y.-H.,  {Manchado} A.,   {Kwitter} K.~B.,  2003,
  \mn@doi [\aj] {10.1086/375206}, \href
  {https://ui.adsabs.harvard.edu/abs/2003AJ....125.3213G} {125, 3213}

\bibitem[\protect\citeauthoryear{{Hamacher}}{{Hamacher}}{2018}]{hamacher2018}
{Hamacher} D.~W.,  2018, \mn@doi [The Australian Journal of Anthropology]
  {10.1111/taja.12257}, \href
  {https://ui.adsabs.harvard.edu/abs/2018AuJAn..29...89H} {29, 89}

\bibitem[\protect\citeauthoryear{{Hamacher} \& {Frew}}{{Hamacher} \&
  {Frew}}{2010}]{hamacher2010}
{Hamacher} D.~W.,  {Frew} D.~J.,  2010, Journal of Astronomical History and
  Heritage, \href {https://ui.adsabs.harvard.edu/abs/2010JAHH...13..220H} {13,
  220}

\bibitem[\protect\citeauthoryear{Hertzog}{Hertzog}{1986}]{hertzog1986}
Hertzog K.~P.,  1986, The Observatory, 106, 38

\bibitem[\protect\citeauthoryear{{Herwig}, {Freytag}  \& {Werner}}{{Herwig}
  et~al.}{2006}]{falk2006}
{Herwig} F.,  {Freytag} B.,   {Werner} K.,  2006, in {Barlow} M.~J.,
  {M{\'e}ndez} R.~H.,  eds,  IAU Symposium Vol. 234, Planetary Nebulae in our
  Galaxy and Beyond. pp 103--110 (\mn@eprint {arXiv} {astro-ph/0606603}),
  \mn@doi{10.1017/S1743921306002833}

\bibitem[\protect\citeauthoryear{Ho}{Ho}{1962}]{ho}
Ho P.~Y.,  1962, Vistas in Astronomy, 5

\bibitem[\protect\citeauthoryear{{Hobbs}, {Lyne}, {Kramer}, {Martin}  \&
  {Jordan}}{{Hobbs} et~al.}{2004}]{hobbs2004}
{Hobbs} G.,  {Lyne} A.~G.,  {Kramer} M.,  {Martin} C.~E.,   {Jordan} C.,  2004,
  \mn@doi [\mnras] {10.1111/j.1365-2966.2004.08157.x}, \href
  {https://ui.adsabs.harvard.edu/abs/2004MNRAS.353.1311H} {353, 1311}

\bibitem[\protect\citeauthoryear{{Hobbs}, {Lorimer}, {Lyne}  \&
  {Kramer}}{{Hobbs} et~al.}{2005}]{hobbs2005}
{Hobbs} G.,  {Lorimer} D.~R.,  {Lyne} A.~G.,   {Kramer} M.,  2005, \mn@doi
  [\mnras] {10.1111/j.1365-2966.2005.09087.x}, \href
  {https://ui.adsabs.harvard.edu/abs/2005MNRAS.360..974H} {360, 974}

\bibitem[\protect\citeauthoryear{Hsi}{Hsi}{1957}]{hsi}
Hsi T.-T.,  1957, Smithonian Contributions to Astrophysics, 2

\bibitem[\protect\citeauthoryear{{Hudec}, {Ba{\v{s}}ta}, {Pihajoki}  \&
  {Valtonen}}{{Hudec} et~al.}{2013}]{hudec2013}
{Hudec} R.,  {Ba{\v{s}}ta} M.,  {Pihajoki} P.,   {Valtonen} M.,  2013, \mn@doi
  [\aap] {10.1051/0004-6361/201219323}, \href
  {https://ui.adsabs.harvard.edu/abs/2013A&A...559A..20H} {559, A20}

\bibitem[\protect\citeauthoryear{{Hulse} \& {Taylor}}{{Hulse} \&
  {Taylor}}{1974}]{hulse1974}
{Hulse} R.~A.,  {Taylor} J.~H.,  1974, \mn@doi [\apjl] {10.1086/181548}, \href
  {https://ui.adsabs.harvard.edu/abs/1974ApJ...191L..59H} {191, L59}

\bibitem[\protect\citeauthoryear{Jacob, Schönberner  \& Steffen}{Jacob
  et~al.}{2013}]{jacob2013}
Jacob R.,  Schönberner D.,   Steffen M.,  2013, A+A, 558, A78

\bibitem[\protect\citeauthoryear{{Kami{\'n}ski}, {Menten}, {Tylenda}, {Hajduk},
  {Patel}  \& {Kraus}}{{Kami{\'n}ski} et~al.}{2015}]{kaminski2015}
{Kami{\'n}ski} T.,  {Menten} K.~M.,  {Tylenda} R.,  {Hajduk} M.,  {Patel}
  N.~A.,   {Kraus} A.,  2015, \mn@doi [\nat] {10.1038/nature14257}, \href
  {https://ui.adsabs.harvard.edu/abs/2015Natur.520..322K} {520, 322}

\bibitem[\protect\citeauthoryear{{Katsuda}, {Tanaka}, {Morokuma}, {Fesen}  \&
  {Milisavljevic}}{{Katsuda} et~al.}{2016}]{katsuda2016}
{Katsuda} S.,  {Tanaka} M.,  {Morokuma} T.,  {Fesen} R.,   {Milisavljevic} D.,
  2016, \mn@doi [\apj] {10.3847/0004-637X/826/2/108}, \href
  {https://ui.adsabs.harvard.edu/abs/2016ApJ...826..108K} {826, 108}

\bibitem[\protect\citeauthoryear{{Kohoutek}}{{Kohoutek}}{2001}]{kohoutek2001}
{Kohoutek} L.,  2001, \mn@doi [\aap] {10.1051/0004-6361:20011162}, \href
  {https://ui.adsabs.harvard.edu/abs/2001A&A...378..843K} {378, 843}

\bibitem[\protect\citeauthoryear{{Kothes}}{{Kothes}}{2013}]{kothes2013}
{Kothes} R.,  2013, \mn@doi [\aap] {10.1051/0004-6361/201219839}, \href
  {https://ui.adsabs.harvard.edu/abs/2013A&A...560A..18K} {560, A18}

\bibitem[\protect\citeauthoryear{{Kwok}}{{Kwok}}{2007}]{kwok2007}
{Kwok} S.,  2007, {The Origin and Evolution of Planetary Nebulae}

\bibitem[\protect\citeauthoryear{{Lazaridis} et~al.,}{{Lazaridis}
  et~al.}{2011}]{lazaridis2011}
{Lazaridis} K.,  et~al., 2011, \mn@doi [\mnras]
  {10.1111/j.1365-2966.2011.18610.x}, \href
  {https://ui.adsabs.harvard.edu/abs/2011MNRAS.414.3134L} {414, 3134}

\bibitem[\protect\citeauthoryear{{Leahy} \& {Tian}}{{Leahy} \&
  {Tian}}{2007}]{leahy2007}
{Leahy} D.~A.,  {Tian} W.~W.,  2007, \mn@doi [\aap]
  {10.1051/0004-6361:20065895}, \href
  {https://ui.adsabs.harvard.edu/abs/2007A&A...461.1013L} {461, 1013}

\bibitem[\protect\citeauthoryear{{Levin} et~al.,}{{Levin}
  et~al.}{2013}]{levin2013}
{Levin} L.,  et~al., 2013, \mn@doi [\mnras] {10.1093/mnras/stt1103}, \href
  {https://ui.adsabs.harvard.edu/abs/2013MNRAS.434.1387L} {434, 1387}

\bibitem[\protect\citeauthoryear{{Manchado}, {Guerrero}, {Kwitter}  \&
  {Chu}}{{Manchado} et~al.}{1992}]{manchado1992}
{Manchado} A.,  {Guerrero} M.,  {Kwitter} K.~B.,   {Chu} Y.~H.,  1992, in
  American Astronomical Society Meeting Abstracts. p. 67.04

\bibitem[\protect\citeauthoryear{{Manchester}, {Hobbs}, {Teoh}  \&
  {Hobbs}}{{Manchester} et~al.}{2005}]{atnf}
{Manchester} R.~N.,  {Hobbs} G.~B.,  {Teoh} A.,   {Hobbs} M.,  2005, \mn@doi
  [\aj] {10.1086/428488}, \href
  {https://ui.adsabs.harvard.edu/abs/2005AJ....129.1993M} {129, 1993}

\bibitem[\protect\citeauthoryear{{Massaro}, {Giommi}, {Leto}, {Marchegiani},
  {Maselli}, {Perri}, {Piranomonte}  \& {Sclavi}}{{Massaro}
  et~al.}{2009}]{massaro2009}
{Massaro} E.,  {Giommi} P.,  {Leto} C.,  {Marchegiani} P.,  {Maselli} A.,
  {Perri} M.,  {Piranomonte} S.,   {Sclavi} S.,  2009, \mn@doi [\aap]
  {10.1051/0004-6361:200810161}, \href
  {https://ui.adsabs.harvard.edu/abs/2009A&A...495..691M} {495, 691}

\bibitem[\protect\citeauthoryear{Mayall}{Mayall}{1967}]{mayall1967}
Mayall M.,  1967, JRASC, 61, 349

\bibitem[\protect\citeauthoryear{Mayall \& Oort}{Mayall \&
  Oort}{1942}]{mayall+oort}
Mayall N.~U.,  Oort J.~H.,  1942, Publications of the Astronomical Society of
  the Pacific, 54, 318

\bibitem[\protect\citeauthoryear{{McCarthy}, {Mendez}  \&
  {Kudritzki}}{{McCarthy} et~al.}{1997}]{mccarthy1997}
{McCarthy} J.~K.,  {Mendez} R.~H.,   {Kudritzki} R.~P.,  1997, in {Habing}
  H.~J.,  {Lamers} H.~J.~G.~L.~M.,  eds,  IAU Symposium Vol. 180, Planetary
  Nebulae. p.~120

\bibitem[\protect\citeauthoryear{{Miller Bertolami}, {Althaus}, {Olano}  \&
  {Jim{\'e}nez}}{{Miller Bertolami} et~al.}{2011a}]{millerBertolami2011}
{Miller Bertolami} M.~M.,  {Althaus} L.~G.,  {Olano} C.,   {Jim{\'e}nez} N.,
  2011a, \mn@doi [\mnras] {10.1111/j.1365-2966.2011.18790.x}, \href
  {https://ui.adsabs.harvard.edu/abs/2011MNRAS.415.1396M} {415, 1396}

\bibitem[\protect\citeauthoryear{{Miller Bertolami}, {Rohrmann}, {Granada}  \&
  {Althaus}}{{Miller Bertolami} et~al.}{2011b}]{millerB2011}
{Miller Bertolami} M.~M.,  {Rohrmann} R.~D.,  {Granada} A.,   {Althaus} L.~G.,
  2011b, \mn@doi [\apjl] {10.1088/2041-8205/743/2/L33}, \href
  {https://ui.adsabs.harvard.edu/abs/2011ApJ...743L..33M} {743, L33}

\bibitem[\protect\citeauthoryear{{Miszalski}, {Parker}, {Acker}, {Birkby},
  {Frew}  \& {Kovacevic}}{{Miszalski} et~al.}{2008}]{mashII2008}
{Miszalski} B.,  {Parker} Q.~A.,  {Acker} A.,  {Birkby} J.~L.,  {Frew} D.~J.,
  {Kovacevic} A.,  2008, \mn@doi [\mnras] {10.1111/j.1365-2966.2007.12727.x},
  \href {https://ui.adsabs.harvard.edu/abs/2008MNRAS.384..525M} {384, 525}

\bibitem[\protect\citeauthoryear{Miszalski, Woudt, Littlefair, P., Warner  \&
  Boffin}{Miszalski et~al.}{2016}]{miszalski2016}
Miszalski B.,  Woudt P.~A.,  Littlefair P. S.,  Warner B.,   Boffin H. M.~J.,
  2016, MNRAS, 456, 633

\bibitem[\protect\citeauthoryear{Morgan}{Morgan}{2007}]{morgan2007}
Morgan J.~A.,  2007, Astrophysics

\bibitem[\protect\citeauthoryear{{Mr{\'o}z} et~al.,}{{Mr{\'o}z}
  et~al.}{2016}]{mroz2016}
{Mr{\'o}z} P.,  et~al., 2016, \mn@doi [\nat] {10.1038/nature19066}, \href
  {https://ui.adsabs.harvard.edu/abs/2016Natur.537..649M} {537, 649}

\bibitem[\protect\citeauthoryear{{Munari}}{{Munari}}{2019}]{munari2019}
{Munari} U.,  2019, arXiv e-prints, \href
  {https://ui.adsabs.harvard.edu/abs/2019arXiv190901389M} {p. arXiv:1909.01389}

\bibitem[\protect\citeauthoryear{{Napiwotzki} \& {Schoenberner}}{{Napiwotzki}
  \& {Schoenberner}}{1995}]{napiwotzki1995}
{Napiwotzki} R.,  {Schoenberner} D.,  1995, \aap, \href
  {https://ui.adsabs.harvard.edu/abs/1995A&A...301..545N} {301, 545}

\bibitem[\protect\citeauthoryear{Neuhäuser, Kunitzsch, Mugrauer, Luge  \& van
  Gent}{Neuhäuser et~al.}{2018}]{neuhaeuserKomet}
Neuhäuser R.,  Kunitzsch P.,  Mugrauer M.,  Luge D.,   van Gent R.,  2018,
  Tycho Brahe, Abu Ma'sar, und der Komet hinter der Venus.
tredition, pp 161--197

\bibitem[\protect\citeauthoryear{{Nice}}{{Nice}}{1995}]{nice1995}
{Nice} D.~J.,  1995, {Pulsar Searches at Arecibo}.
p.~9

\bibitem[\protect\citeauthoryear{{Ohsawa}, {Onaka}, {Sakon}, {Mori}, {Miyata},
  {Asano}, {Matsuura}  \& {Kaneda}}{{Ohsawa} et~al.}{2012}]{ohsawa2012}
{Ohsawa} R.,  {Onaka} T.,  {Sakon} I.,  {Mori} T.~I.,  {Miyata} T.,  {Asano}
  K.,  {Matsuura} M.,   {Kaneda} H.,  2012, \mn@doi [\apjl]
  {10.1088/2041-8205/760/2/L34}, \href
  {https://ui.adsabs.harvard.edu/abs/2012ApJ...760L..34O} {760, L34}

\bibitem[\protect\citeauthoryear{{Parker} et~al.,}{{Parker}
  et~al.}{2006}]{mash2006}
{Parker} Q.~A.,  et~al., 2006, \mn@doi [\mnras]
  {10.1111/j.1365-2966.2006.10950.x}, \href
  {https://ui.adsabs.harvard.edu/abs/2006MNRAS.373...79P} {373, 79}

\bibitem[\protect\citeauthoryear{Patterson et~al.,}{Patterson
  et~al.}{2013}]{pat2013}
Patterson J.,  et~al., 2013, MNRAS, 434, 1902

\bibitem[\protect\citeauthoryear{{Pe{\~n}a}, {Ruiz-Escobedo},
  {Rechy-Garc{\'\i}a}  \& {Garc{\'\i}a-Rojas}}{{Pe{\~n}a}
  et~al.}{2017}]{pena2017}
{Pe{\~n}a} M.,  {Ruiz-Escobedo} F.,  {Rechy-Garc{\'\i}a} J.,
  {Garc{\'\i}a-Rojas} J.,  2017, in {Liu} X.,  {Stanghellini} L.,   {Karakas}
  A.,  eds,  IAU Symposium Vol. 323, Planetary Nebulae: Multi-Wavelength Probes
  of Stellar and Galactic Evolution. pp 60--64 (\mn@eprint {arXiv}
  {1611.06198}), \mn@doi{10.1017/S1743921317000850}

\bibitem[\protect\citeauthoryear{{Perek} \& {Kohoutek}}{{Perek} \&
  {Kohoutek}}{1967}]{perek1967}
{Perek} L.,  {Kohoutek} L.,  1967, {Catalogue of Galactic Planetary Nebulae}

\bibitem[\protect\citeauthoryear{{Pereyra}, {Richer}  \& {L{\'o}pez}}{{Pereyra}
  et~al.}{2013}]{pereyra2013}
{Pereyra} M.,  {Richer} M.~G.,   {L{\'o}pez} J.~A.,  2013, \mn@doi [\apj]
  {10.1088/0004-637X/771/2/114}, \href
  {https://ui.adsabs.harvard.edu/abs/2013ApJ...771..114P} {771, 114}

\bibitem[\protect\citeauthoryear{Protte \& Hoffmann}{Protte \&
  Hoffmann}{2020}]{protteHoffmann2020}
Protte P.,  Hoffmann S.~M.,  {forseen for 2020}, Astron. Notes

\bibitem[\protect\citeauthoryear{Pskovskii}{Pskovskii}{1972}]{pskovskii}
Pskovskii Y.~P.,  1972, Soviet Astronomy, 16

\bibitem[\protect\citeauthoryear{{Ray}, {Thorsett}, {Jenet}, {van Kerkwijk},
  {Kulkarni}, {Prince}, {Sand hu}  \& {Nice}}{{Ray} et~al.}{1996}]{ray1996}
{Ray} P.~S.,  {Thorsett} S.~E.,  {Jenet} F.~A.,  {van Kerkwijk} M.~H.,
  {Kulkarni} S.~R.,  {Prince} T.~A.,  {Sand hu} J.~S.,   {Nice} D.~J.,  1996,
  \mn@doi [\apj] {10.1086/177934}, \href
  {https://ui.adsabs.harvard.edu/abs/1996ApJ...470.1103R} {470, 1103}

\bibitem[\protect\citeauthoryear{{Reynolds} \& {Borkowski}}{{Reynolds} \&
  {Borkowski}}{2019}]{reynolds2019}
{Reynolds} S.~P.,  {Borkowski} K.~J.,  2019, \mn@doi [\apj]
  {10.3847/1538-4357/ab5804}, \href
  {https://ui.adsabs.harvard.edu/abs/2019ApJ...887..233R} {887, 233}

\bibitem[\protect\citeauthoryear{{Rivinius}, {Baade}, {Hadrava}, {Heida}  \&
  {Klement}}{{Rivinius} et~al.}{2020}]{rivinius2020}
{Rivinius} T.,  {Baade} D.,  {Hadrava} P.,  {Heida} M.,   {Klement} R.,  2020,
  \mn@doi [\aap] {10.1051/0004-6361/202038020}, \href
  {https://ui.adsabs.harvard.edu/abs/2020A&A...637L...3R} {637, L3}

\bibitem[\protect\citeauthoryear{{Rosado}}{{Rosado}}{1982}]{rosado1982}
{Rosado} M.,  1982, \rmxaa, \href
  {https://ui.adsabs.harvard.edu/abs/1982RMxAA...5..127R} {5, 127}

\bibitem[\protect\citeauthoryear{{Sabbadin}, {Bianchini}, {Ortolani}  \&
  {Strafella}}{{Sabbadin} et~al.}{1985}]{sabbadin1985}
{Sabbadin} F.,  {Bianchini} A.,  {Ortolani} S.,   {Strafella} F.,  1985,
  \mn@doi [\mnras] {10.1093/mnras/217.3.539}, \href
  {https://ui.adsabs.harvard.edu/abs/1985MNRAS.217..539S} {217, 539}

\bibitem[\protect\citeauthoryear{{Schaefer}, {Pagnotta}  \& {Shara}}{{Schaefer}
  et~al.}{2010}]{schaefer2010}
{Schaefer} B.~E.,  {Pagnotta} A.,   {Shara} M.~M.,  2010, \mn@doi [\apj]
  {10.1088/0004-637X/708/1/381}, \href
  {https://ui.adsabs.harvard.edu/abs/2010ApJ...708..381S} {708, 381}

\bibitem[\protect\citeauthoryear{Schlier}{Schlier}{1935}]{schlier}
Schlier O.,  1935, AN, 254, 181

\bibitem[\protect\citeauthoryear{{Shappee} et~al.,}{{Shappee}
  et~al.}{2014}]{shappee2014}
{Shappee} B.~J.,  et~al., 2014, \mn@doi [\apj] {10.1088/0004-637X/788/1/48},
  \href {https://ui.adsabs.harvard.edu/abs/2014ApJ...788...48S} {788, 48}

\bibitem[\protect\citeauthoryear{{Shara}, {Moffat}, {McGraw}, {Dearborn},
  {Bond}, {Kemper}  \& {Lamontagne}}{{Shara} et~al.}{1984}]{shara1984}
{Shara} M.~M.,  {Moffat} A.~F.~J.,  {McGraw} J.~T.,  {Dearborn} D.~S.,  {Bond}
  H.~E.,  {Kemper} E.,   {Lamontagne} R.,  1984, \mn@doi [\apj]
  {10.1086/162260}, \href
  {https://ui.adsabs.harvard.edu/abs/1984ApJ...282..763S} {282, 763}

\bibitem[\protect\citeauthoryear{Shara, Drissen, Martin, Alarie  \&
  Stephenson}{Shara et~al.}{2017a}]{shara2017_ATcnc_steph}
Shara M.~M.,  Drissen L.,  Martin T.,  Alarie A.,   Stephenson F.~R.,  2017a,
  MNRAS, 465, 739

\bibitem[\protect\citeauthoryear{Shara et~al.,}{Shara
  et~al.}{2017b}]{shara2017_nov1437}
Shara M.~M.,  et~al., 2017b, Nature, 548, 558

\bibitem[\protect\citeauthoryear{{Sowicka}, {Jones}, {Corradi}, {Wesson},
  {Garc{\'\i}a-Rojas}, {Santand er-Garc{\'\i}a}, {Boffin}  \&
  {Rodr{\'\i}guez-Gil}}{{Sowicka} et~al.}{2017}]{sowicka2017}
{Sowicka} P.,  {Jones} D.,  {Corradi} R. L.~M.,  {Wesson} R.,
  {Garc{\'\i}a-Rojas} J.,  {Santand er-Garc{\'\i}a} M.,  {Boffin} H. M.~J.,
  {Rodr{\'\i}guez-Gil} P.,  2017, \mn@doi [\mnras] {10.1093/mnras/stx1697},
  \href {https://ui.adsabs.harvard.edu/abs/2017MNRAS.471.3529S} {471, 3529}

\bibitem[\protect\citeauthoryear{{Stafford} \& {Lopez}}{{Stafford} \&
  {Lopez}}{2020}]{stafford2020}
{Stafford} J.~N.,  {Lopez} L.,  2020, in American Astronomical Society Meeting
  Abstracts. American Astronomical Society Meeting Abstracts.
p. 307.09

\bibitem[\protect\citeauthoryear{{Stanghellini}, {Garc{\'\i}a-Hern{\'a}ndez},
  {Garc{\'\i}a-Lario}, {Davies}, {Shaw}, {Villaver}, {Manchado}  \&
  {Perea-Calder{\'o}n}}{{Stanghellini} et~al.}{2012}]{stanghellini2012}
{Stanghellini} L.,  {Garc{\'\i}a-Hern{\'a}ndez} D.~A.,  {Garc{\'\i}a-Lario} P.,
   {Davies} J.~E.,  {Shaw} R.~A.,  {Villaver} E.,  {Manchado} A.,
  {Perea-Calder{\'o}n} J.~V.,  2012, \mn@doi [\apj]
  {10.1088/0004-637X/753/2/172}, \href
  {https://ui.adsabs.harvard.edu/abs/2012ApJ...753..172S} {753, 172}

\bibitem[\protect\citeauthoryear{{Stanghellini}, {Shaw}  \&
  {Villaver}}{{Stanghellini} et~al.}{2016}]{stanghellini2016}
{Stanghellini} L.,  {Shaw} R.~A.,   {Villaver} E.,  2016, \mn@doi [\apj]
  {10.3847/0004-637X/830/1/33}, \href
  {https://ui.adsabs.harvard.edu/abs/2016ApJ...830...33S} {830, 33}

\bibitem[\protect\citeauthoryear{Stephenson}{Stephenson}{1976}]{stephenson}
Stephenson F.~R.,  1976, Quarterly Journal Royal Astronomical Society, 17, 121

\bibitem[\protect\citeauthoryear{Stephenson \& Green}{Stephenson \&
  Green}{2002}]{stephGreen}
Stephenson F.~R.,  Green D.~A.,  2002, Historical Supernovae and their
  remnants.
Oxford U Press, New York

\bibitem[\protect\citeauthoryear{Tappert, Ederoclite, Mennickent,
  Schmidtobreick  \& Vogt}{Tappert et~al.}{2012}]{tappert2012}
Tappert C.,  Ederoclite A.,  Mennickent R.~E.,  Schmidtobreick L.,   Vogt N.,
  2012, \mn@doi [MNRAS] {10.1111/j.1365-2966.2012.21054.x}, 423, 2476

\bibitem[\protect\citeauthoryear{{Thorsett}, {Deich}, {Kulkarni}, {Navarro}  \&
  {Vasisht}}{{Thorsett} et~al.}{1993}]{thorsett1993}
{Thorsett} S.~E.,  {Deich} W.~T.~S.,  {Kulkarni} S.~R.,  {Navarro} J.,
  {Vasisht} G.,  1993, \mn@doi [\apj] {10.1086/173224}, \href
  {https://ui.adsabs.harvard.edu/abs/1993ApJ...416..182T} {416, 182}

\bibitem[\protect\citeauthoryear{{Todt} et~al.,}{{Todt}
  et~al.}{2015}]{todt2015}
{Todt} H.,  et~al., 2015, {The Born-again Planetary Nebulae Abell 30 and Abell
  78}.
p.~141

\bibitem[\protect\citeauthoryear{{Tweedy} \& {Kwitter}}{{Tweedy} \&
  {Kwitter}}{1996}]{tweedy1996}
{Tweedy} R.~W.,  {Kwitter} K.~B.,  1996, \mn@doi [\apjs] {10.1086/192364},
  \href {https://ui.adsabs.harvard.edu/abs/1996ApJS..107..255T} {107, 255}

\bibitem[\protect\citeauthoryear{Warner}{Warner}{1995}]{warner1995}
Warner B.,  1995, Cataclysmic Variable Stars.
Cambridge Astrophysical Series, Cambridge University Press

\bibitem[\protect\citeauthoryear{{Watson}, {Henden}  \& {Price}}{{Watson}
  et~al.}{2006}]{watson}
{Watson} C.~L.,  {Henden} A.~A.,   {Price} A.,  2006, Society for Astronomical
  Sciences Annual Symposium, \href
  {https://ui.adsabs.harvard.edu/abs/2006SASS...25...47W} {25, 47}

\bibitem[\protect\citeauthoryear{{Wenger} et~al.,}{{Wenger}
  et~al.}{2000}]{wenger2000}
{Wenger} M.,  et~al., 2000, \mn@doi [\aaps] {10.1051/aas:2000332}, \href
  {https://ui.adsabs.harvard.edu/abs/2000A&AS..143....9W} {143, 9}

\bibitem[\protect\citeauthoryear{{Woudt} \& {Warner}}{{Woudt} \&
  {Warner}}{2002}]{woudt2002}
{Woudt} P.~A.,  {Warner} B.,  2002, \mn@doi [\mnras]
  {10.1046/j.1365-8711.2002.05613.x}, \href
  {https://ui.adsabs.harvard.edu/abs/2002MNRAS.335...44W} {335, 44}

\bibitem[\protect\citeauthoryear{Xi \& Po}{Xi \& Po}{1966}]{xi+po}
Xi Z.-z.,  Po S.-j.,  1966, Science, 154

\bibitem[\protect\citeauthoryear{Xu, Pankenier  \& Jiang}{Xu
  et~al.}{2000}]{xu2000}
Xu Z.,  Pankenier D.~W.,   Jiang Y.,  2000, East Asian Archaeoastronomy.
Gordon and Breach Science Publishers

\bibitem[\protect\citeauthoryear{Yau}{Yau}{1988}]{yau}
Yau K. K.~C.,  1988, PhD thesis, Durham University, \url
  {http://etheses.dur.ac.uk/6331/1/6331_3685.PDF?UkUDh:CyT}

\bibitem[\protect\citeauthoryear{Zotti \& Wolf}{Zotti \&
  Wolf}{2020}]{stellarium}
Zotti G.,  Wolf A.,  2020, Stellarium User Guide.
\url {http://stellarium.org}

\bibitem[\protect\citeauthoryear{{della Valle}, {Gilmozzi}, {Bianchini}  \&
  {Esenoglu}}{{della Valle} et~al.}{1997}]{valle1997}
{della Valle} M.,  {Gilmozzi} R.,  {Bianchini} A.,   {Esenoglu} H.,  1997,
  \aap, \href {https://ui.adsabs.harvard.edu/abs/1997A&A...325.1151D} {325,
  1151}

\makeatother
\end{thebibliography}


\begin{thebibliography}{99}
\bibitem[\protect\citeauthoryear{Paper\,1}{}]{vogt2019} 
 \textbf{Paper\,1:} Vogt, Nikolaus and Hoffmann, Susanne M. and Tappert, Claus, On the possibilities of classical nova identifications among historical Far Eastern guest star observations, Astronomische Nachrichten, 2019, 752, 340
\bibitem[\protect\citeauthoryear{Paper\,2}{}]{hoffmann2019}
 \textbf{Paper\,2:} Hoffmann, Susanne, What information can we derive from historical Far Eastern guest stars for modern research on novae and cataclysmic variables?, MNRAS, 2019, 490, 4194 
\bibitem[\protect\citeauthoryear{Paper\,3}{}]{hoffmannVogtProtte}
 \textbf{Paper\,3:} Hoffmann, S. M. and Vogt, N. and Protte, P., A new approach to generate a catalogue of potential historical novae, Astronomische Nachrichten, 2020, 341, 79
\bibitem[\protect\citeauthoryear{Paper\,4}{}]{hovoMNRAS2020}
 \textbf{Paper\,4:} Hoffmann, Susanne M. and  Vogt, Nikolaus, Cataclysmic variables as possible counterparts of ancient Far Eastern guest stars, MNRAS, 2020, 494, 5775
\bibitem[\protect\citeauthoryear{Paper\,5}{}]{hoffmannVogtLux}
\end{thebibliography}




\appendix
\definecolor{darkred}{rgb}{.7,0,0} 
\definecolor{pink}{rgb}{.9,.7,.7} 

\section{Online Only Appendix} 
 The star charts show the details of the distribution of objects in the considered fields. Thus, we provide one PDF-file which include overview maps. However, as all object densities are too high to label the plotted objects individually, we also provide the same maps as CDF files in a ZIP folder. 
 
\subsection{Plots of Planetary Nebulae (and PN candidates)}
This appendix displays for each historical event two maps syn-optically. On the left hand side, we plotted our search field and all cataclysmic variables (CV), all planetary nebulae (PN) and PN candidates listed in Simbad into our map. The PN and PN candidates are plotted in dark/light red colour, respectively, the CVs and low mass X-ray binaries are plotted in green/ cyan. On the right hand side, next to this chart, we display the same map with our search field but displayed only those nebulae (PN, PN candidates) which have a nearby (in 2D-projection) close binary. In these maps, the nebulae are mapped in blue while their nearby binaries are mapped in orange (symbiotic stars), cyan (X-ray binaries), and green (CVs). In both maps planetary nebulae (PN) are indicated by a ring with central point $\odot$, PN candidates with $\otimes$, and all types of binary stars with $\diamondsuit$. The catalogue of PN and PN candidates originates from Simbad, downloaded in January 2020, while the catalogues of CV, LMXB, and Symbiotic stars are downloaded from the VSX catalogue of the AAVSO (January to March 2020). 
\begin{figure}
	\includegraphics[width=\columnwidth]{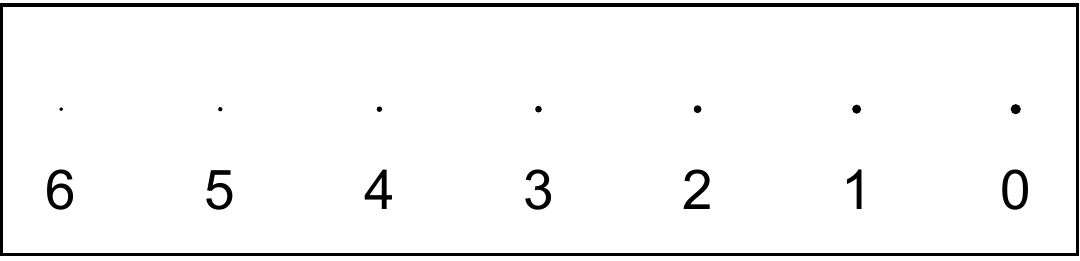} 
    \caption{Scale of star sizes for the maps in the appendix.}
    \label{fig:magScale}
\end{figure}

All maps include the Chinese asterism lines and the stars scaled according to their magnitude (Fig.~\ref{fig:magScale}).


\bsp	
\label{lastpage}
\end{document}